\documentclass[%
aip,jcp,amsmath,amssymb,
preprint,%
citeautoscript,
]{revtex4-2}

\usepackage[utf8]{inputenc}

\usepackage[dvipsnames]{xcolor}
\usepackage{tikz}
\usetikzlibrary{shapes}
\DeclareRobustCommand{\A}{\raisebox{0.4pt}{\tikz{\node[draw,scale=0.48,circle,fill=black](){};}}}
\DeclareRobustCommand{\B}{\raisebox{0.4pt}{\tikz{\node[draw,scale=0.48,circle,](){};}}}

\usepackage{graphicx}

\begin{document}

\preprint{--}

\title{Unsupervised learning of sequence-specific aggregation behavior for a model copolymer}
\author{Antonia Statt}
\affiliation{Materials Science and Engineering, Grainger College of Engineering, University of Illinois, Urbana-Champaign, IL 61801 }

\author{Devon C. Kleeblatt}
\affiliation{Materials Science and Engineering, Pennsylvania State University, University Park, PA 16802}

\author{Wesley F. Reinhart}
\email[email:]{reinhart@psu.edu}
\affiliation{Materials Science and Engineering, Pennsylvania State University, University Park, PA 16802}
\affiliation{Institute for Computational and Data Sciences, Pennsylvania State University, University Park, PA 16802}

\date{27 July 2021}

\begin{abstract}
We apply a recently developed unsupervised machine learning scheme for local atomic environments ~\cite{Reinhart2021} to characterize large-scale, disordered aggregates formed by sequence-defined macromolecules.
This method provides new insight into the structure of these disordered, dilute aggregates, which has proven difficult to understand using collective variables manually derived from expert knowledge~\cite{Statt2020}.
In contrast to such conventional order parameters, we are able to classify the global aggregate structure directly using descriptions of the local environments.
The resulting characterization provides a deeper understanding of the range of possible self-assembled structures and their relationships to each other.
We also provide detailed analysis of the effects of finite system size, stochasticity, and kinetics of these aggregates based on the learned collective variables.
Interestingly, we find that the spatiotemporal evolution of systems in the learned latent space is smooth and continuous, despite being derived from only a single snapshot from each of about $1 \, 000$ monomer sequences.
These results demonstrate the insight which can be gained by applying unsupervised machine learning to soft matter systems, especially when suitable order parameters are not known.
\end{abstract}


\maketitle

\section{Introduction}

A major hurdle for the data driven design of functional soft polymeric materials is the topological and chemical complexity of the constituent macromolecules.
As a result, their material properties, such as large-scale morphologies and aggregation dynamics, are defined by features which span many length and timescales, making computational investigation challenging.
While there are a few recent applications of machine learning for disordered soft matter systems \cite{Webb2020,Jablonka2021,Afzal2019,Shmilovich2020,Wang2020}, the best practices for representation and classification of complex disordered systems remain unclear.

The bulk self-assembly of polymer chains has been studied extensively for various architectures of synthetic polymers,~\cite{Matsen2012,Zhang2017,Bates2017,Levine2016,Mai2012} including multiblock copolymers~\cite{Bates2012,Wu2004} and tapered blocks.~\cite{Pakula1996,Beranek2020} 
Aggregate formation is commonly observed and well characterized in block copolymers, ~\cite{Koch2015, Floriano1999, Li2019, Posocco2010, Dolgov2018} where self-assembly is driven by the microphase separation of the different blocks.
Aggregation behavior has also been extensively studied for dilute systems in solution, specifically with di- and tri-blocks~\cite{Dolgov2018,Li2019,Fenyves2014} and multiblocks,~\cite{Gindy2008} where the focus was mainly to describe finite-size aggregates like spherical micelles and vesicles, and gelation.~\cite{Hugouvieux2009,Hugouvieux2011}

Perhaps the most common example of sequence-defined macromolecules are proteins.
The polymers investigated here do not exhibit secondary or tertiary structures, and are more similar to intrinsically disordered proteins (IDP).
In fact, the model used here was first considered in the context of a simple IDP model for liquid-liquid phase separation\cite{Statt2020}.
In this work, we focus on the sequence dependence of phase and aggregation behavior of macromolecules which self-assemble into large-scale, disordered aggregates.

To identify local structures, bond order parameters~\cite{Lechner2008,Steinhardt1983} or graph-based descriptors~\cite{Reinhart2017} are commonly used. 
This approach is not very well suited for disordered or liquid aggregates, because the topological environments of disordered aggregates are random. 
Fully supervised machine learning approaches are also not appropriate, because the number and nature of possible structures in the disordered aggregates are unknown.
While structure factors or scattering functions contain longer range features of the large-scale aggregates, they typically requires fitting the scattering data with theoretical models, which can be open to interpretation.~\cite{AKCASU1980866,PEDERSEN1997171} In simulations, only a finite range of $q$-vectors is accessible, complicating analysis further.

To address these challenges, unsupervised machine learning has become increasingly popular in recent years.
One widespread approach is the use of diffusion maps\cite{Coifman2006} to identify low-dimensional collective variables directly from particle tracking or simulation data \cite{Ferguson2011}, which has been employed to study colloidal clusters \cite{Long2015} and crystals \cite{Reinhart2017}, ring polymers \cite{Wang2018}, and nematic phases \cite{Chiappini2020}, to name only a few.
The development of this approach was rooted in fundamental statistical mechanics \cite{Zwanzig2001}, wherein the existence of a diffusion process is postulated that exhibits some slow dynamical modes coupled to fast stochastic noise.

While soft matter is still analyzed using conventional methods such as principal component analysis \cite{Xu2019}, many sophisticated methods for dimensionality reduction (also called manifold learning) have emerged from the machine learning literature in the last decade.
This progress has not escaped the attention of those interested in soft matter physics, and a variety of those techniques have now been applied to study polymer aggregation.
The conformations of single polymer chains has been studied by a nonmetric multidimensional scaling \cite{Bejagam2018} while aggregates have been analyzed by UMAP \cite{Ziolek2021}, both of which are similar in spirit to diffusion maps.
Self-supervised neural networks (e.g., autoencoders) have also seen impressive success in capturing the conformations of single polymer chains \cite{Chen2018, Sun2020, Bhattacharya2021}.

In this work, we utilize a recently developed unsupervised machine learning method for local environments based on simple rotation-invariant features~\cite{Reinhart2021} to characterize the morphologies of self-assembled aggregates.
This method does not rely on discrete categories for structural identification, and has been shown to be effective with very few training data.
While it has previously been demonstrated on a variety of soft matter systems such as colloidal self-assembly, ice structures, and binary mesophases~\cite{Reinhart2021}, this is the first demonstration of that method on macromolecules.
As in other recent works involving unsupervised learning for soft matter, we use nonlinear manifold learning to discover collective variables which describe the aggregate morphologies and assembly dynamics.
However, we consider larger aggregates with more structural variation compared to recent works (e.g., Refs.~\citenum{Ziolek2021} and \citenum{Bhattacharya2021}) which in turn must be described by more features.

We focus our attention on exploring the diversity of the disordered aggregates and interpreting the discovered latent space.
Our aims are as follows: (i) find a reliable, unbiased method to classify disordered, dilute, large-scale aggregates, and (ii) apply this method to deepen the physical understanding of the self-assembly of a model sequence-defined macromolecule.

In this article, we first provide a description of each identified aggregate morphology (from over $2 \, 000$ simulated sequences) in Section~\ref{sec:aggregates} based on the learned latent space.
Next we discuss the bias introduced by random sampling of sequences in Section~\ref{sec:sampling}.
We compare the obtained latent space to expert knowledge (i.e., the order parameters from Ref.~\citenum{Statt2020}) in Section~\ref{sec:expert_knowledge}, and investigate how the finite system size (Section~\ref{sec:finite-size}) and stochasticity (Section~\ref{sec:variability}) of our simulations influence the results.
We also provide a detailed analysis of the aggregation kinetics with respect to the latent space collective variables in Section~\ref{sec:trajectories}, showing that the manifold learned from individual sequences is continuous with respect to a single trajectory.

\section{Methods \label{sec:methods}}

\subsection{Molecular Dynamics simulations}

To investigate complex aggregation behavior of soft matter, we used a very simple model with a few parameters, which shows a rich self-assembly behavior. 
Although the model was first used to investigate disordered protein liquid-liquid phase separation,~\cite{Statt2020} it is a very general model of a sequence-defined macromolecule, like a random block copolymer.
Each chain consists of a sequence of sticky $A$, and non-sticky $B$ beads.
The sticky beads interact with each other via the standard Lennard-Jones (LJ) potential ($\sigma$, $\epsilon$, $r_\text{cut}=3\sigma$)~\cite{Jones1924}.
The non-sticky self- and cross-interactions were described by the purely repulsive Weeks-Chandler-Anderson (WCA) potential~\cite{Weeks1971}.
Bonds between subsequent beads in the chain are described by the standard finitely extensible nonlinear elastic (FENE) potential ($R_0=1.5\sigma$,$K=30\epsilon/\sigma^2$)~\cite{Kremer1990}. For computational efficiency, we use an implicit-solvent model, and solvent  effects, such as hydrodynamics, are neglected. For a detailed description of the model we refer to Ref.~\citenum{Statt2020}.

In this work, we varied only the sequence of bead types and kept all other parameters constant.
The fraction $f_A$ of $A$ beads was set to $0.6$ and the chain length $M$ was constant at $M=20$ for all investigated sequences, resulting in unentangled chains at a density of $\rho=0.05\, m/\sigma^3$, in a box of size $V\approx(40\sigma)^3$ with $N=500$ chains.
To evaluate finite size effects, we kept the density fixed but simulated $N=(20, 50, 100, 200, 500, 1000, 2000)$ chains.

All simulations were performed using the HOOMD-blue (version 2.9.4) simulation package~\cite{Glaser2015,Anderson2008} on graphics processing units.
We performed $NVT$ simulations with a Langevin thermostat to keep temperature constant at $T=0.5\epsilon$.
Simulation time varied from $5 \times 10^3 \, \tau$ for investigating the short-term evolution to $2 \times 10^5 \, \tau$ for studying long-term equilibration of the aggregates.

\begin{figure}[!t]
    \centering
    \includegraphics[width=90mm]{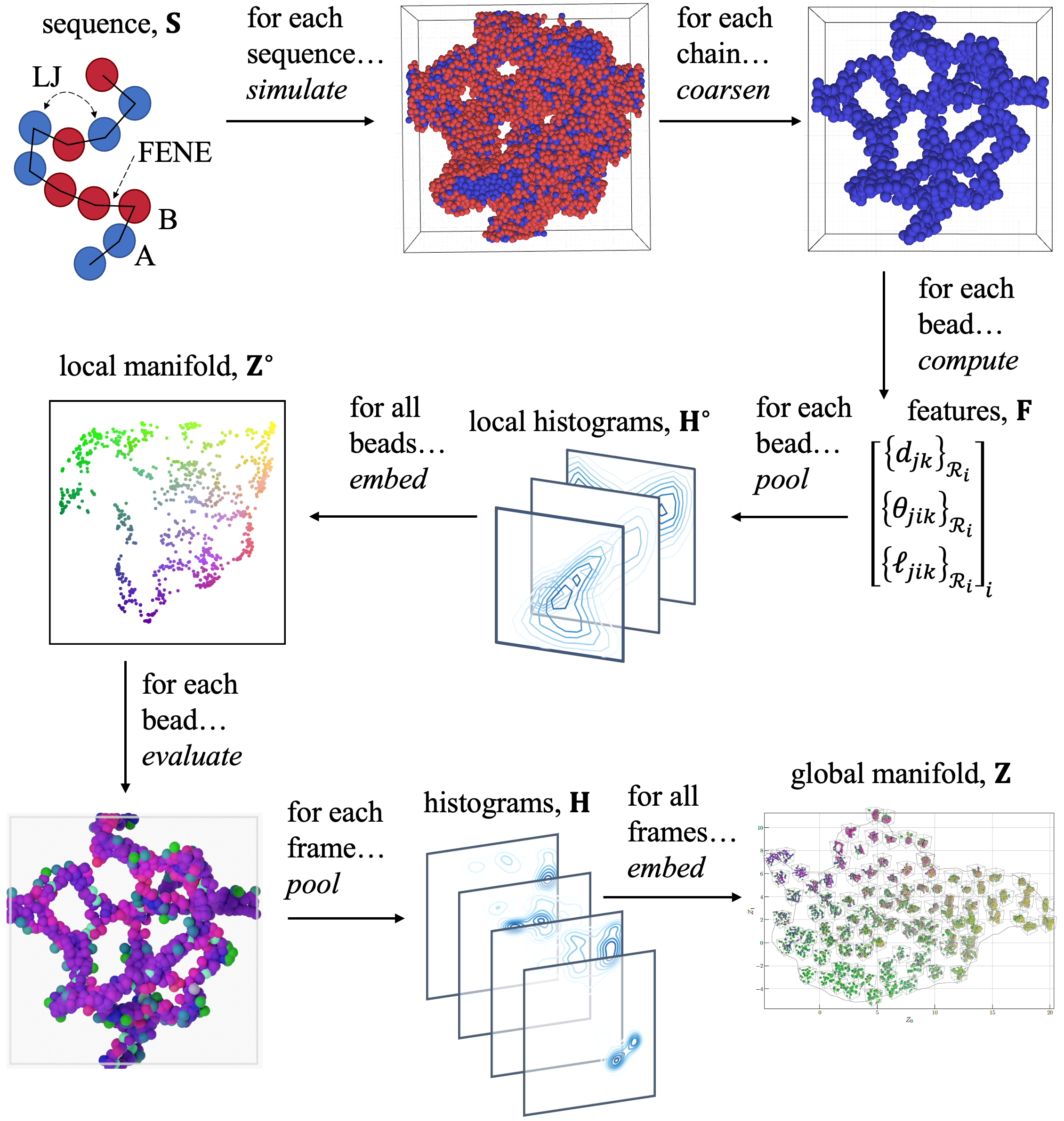}
    \caption{Schematic of the unsupervised learning method used to characterize the local structure ($\mathbf{Z}^\circ$) from local environment feature histograms ($\mathbf{H}^\circ_i$) and global aggregate morphology ($\mathbf{Z}$) from a system wide histogram ($\mathbf{H}$) of particle positions in $\mathbf{Z}^\circ$ space.}
    \label{fig:schematic}
\end{figure}

\subsection{Unsupervised machine learning}

The self-assembled structures are classified based on a recently introduced unsupervised learning approach for local environments \cite{Reinhart2021}.
The goal of this approach is to find a set of order parameters for the large-scale aggregates based on the geometric characteristics of the local environments around each polymer chain.
As implemented here, the geometric characteristics are a collection of three-body terms which are combined into a single feature vector through a histogram-like kernel density estimation.
The resulting local descriptors are reduced to a a few local order parameters using manifold learning and then combined through a second kernel density estimation to create features for the entire simulation snapshot (i.e., characterizing the aggregate).  
Finally, these global features are embedded into a second manifold to characterize the self-assembled structures by only two numbers describing a position in 2D latent space.
A schematic of the procedure is shown in Fig.~\ref{fig:schematic}.

As input for each particle $i$, the method takes a local neighborhood $\mathbf{R}_i$ according to an isotropic cutoff radius, with $n_i$ particles inside this $r_\mathrm{cut} = 6 \, \sigma$ radius.
From these, the three-body features $\mathbf{F}_i$ between particles $(i, j, k)$ are computed:
\begin{itemize}
\item $d_{jk} = |\mathbf{r}_k - \mathbf{r}_j|$ is the distance between neighbors
\item $\theta_{jik} = \arccos \left( \mathbf{r}_{ik} \cdot \mathbf{r}_{ij} \right)$ is the bond angle
\item $\ell_{jik} = d_{ij} + d_{ik}$ is the bond length
\end{itemize}
Here $\mathbf{r}_{ij} = \mathbf{r}_j - \mathbf{r}_i$ is the displacement vector between particle $i$ and $j$.
Since the central particle (i.e., defining the origin of the chain neighborhood) necessarily participates in every bond, there are $3 \, n_i^2$ features.

Once these features are computed for the entire neighborhood, permutation invariance is enforced by performing a kernel density estimation in two dimensions for each of the $(d, \theta)$, $(d, \ell)$, and $(\theta, \ell)$ slices and summing over each bin to yield a histogram $\mathbf{H}^\circ_i$.
Each entry in this matrix represents the relative probability that a given pair of the three-body terms is observed in the local environment.
Here 12 bins are used in each dimension, and the matrix is normalized (i.e., yielding a probability density) to avoid a strong dependence on the number of particles.
Some example $\mathbf{H}^\circ$ histograms are shown in Fig.~S5 of the ESI\cite{SI} to illustrate the difference between local particle environments in different aggregate types.
Note that on their own, these features do not provide a very useful description of the local environments.
Instead, the intent is to generate a large amount of information about each local environment and later distill it down into a reduced form (i.e., an order parameter) based on preservation of the high-dimensional topology in the learned low-dimensional manifold.

This histogram is reshaped into a $3 \, n_\mathrm{bins}^2 \times 1$ feature vector and embedded into a low-dimensional manifold $Z$ using the UMAP algorithm \cite{Mcinnes2020}.
UMAP is a non-linear, unsupervised method for dimensionality reduction that assumes a uniform distribution of data over a locally connected Riemannian manifold.
This results in a low-dimensional projection of the original data (i.e., the collection of local feature histograms $\mathbf{H}^\circ_i$) which approximately maintains its topological structure.
We refer to this projection as the latent space $\mathbf{Z}^\circ$, which contains the information about local particle environments.
The hyperparameters for the UMAP algorithm are 10 neighbors, zero minimum distance, and 3 components in the latent space.
A more complete description of the method can be found in Ref.~\citenum{Reinhart2021}.

Evaluating the particles based on their position in $\mathbf{Z}^\circ$ provides local structural information about each polymer chain (e.g., used to color the particles in Fig.~\ref{fig:structures}) but still does not characterize aggregates.
For instance, the neighborhood around a liquid should be relatively isotropic, whereas a membrane or string neighborhood should be strongly anisotropic.
In order to evaluate the global morphology, the structural information for all particles in the simulation domain need to be considered collectively.
This is achieved by a second pooling and embedding step using the same procedure, as described in Ref.~\citenum{Reinhart2021} to generate collective variables.
A histogram of the positions in $\mathbf{Z}^\circ$ is generated for all particles in the simulation, and this system wide $\mathbf{H}$ is then flattened to create a second feature vector, this time for the entire snapshot.
Some example $\mathbf{H}$ histograms are shown in Fig.~S6 of the ESI\cite{SI}.
This histogram is again flattened and embedded into a second latent space $\mathbf{Z}$ to construct a manifold of all observed morphologies.
This second latent space $\mathbf{Z}$ now contains the low-dimensional data for each simulation snapshot (i.e., a collection of all chains), rather than each particle, and can be used as a collective variable to characterize the aggregates.
As in Ref.~\citenum{Reinhart2021}, the hyperparameters are modified in this second embedding step to yield a smoother manifold, with 16 neighbors, 1 minimum distance, and 2 components in the latent space.

While the method has been demonstrated on different chemical species\cite{Reinhart2021}, here we ignore the monomer type and consider only structural features.
To accelerate the analysis, we also further coarse-grain the chains by considering only two particles per chain, as defined by the center of mass of the first and last ten monomer beads.
Note this does not affect the simulation model and is only performed in post-processing.
In principle this step is not necessary, but the analysis of the full system required considerable computation time -- even with GPU acceleration -- and we determined that the coarse-grained version retained all of the essential morphological features (e.g., the string topology in the top-right of Fig.~\ref{fig:schematic}).

\section{Results\label{sec:results}}

\subsection{Aggregate morphology\label{sec:aggregates}}

\begin{figure}[!t]
    \centering
    \includegraphics[width=8cm]{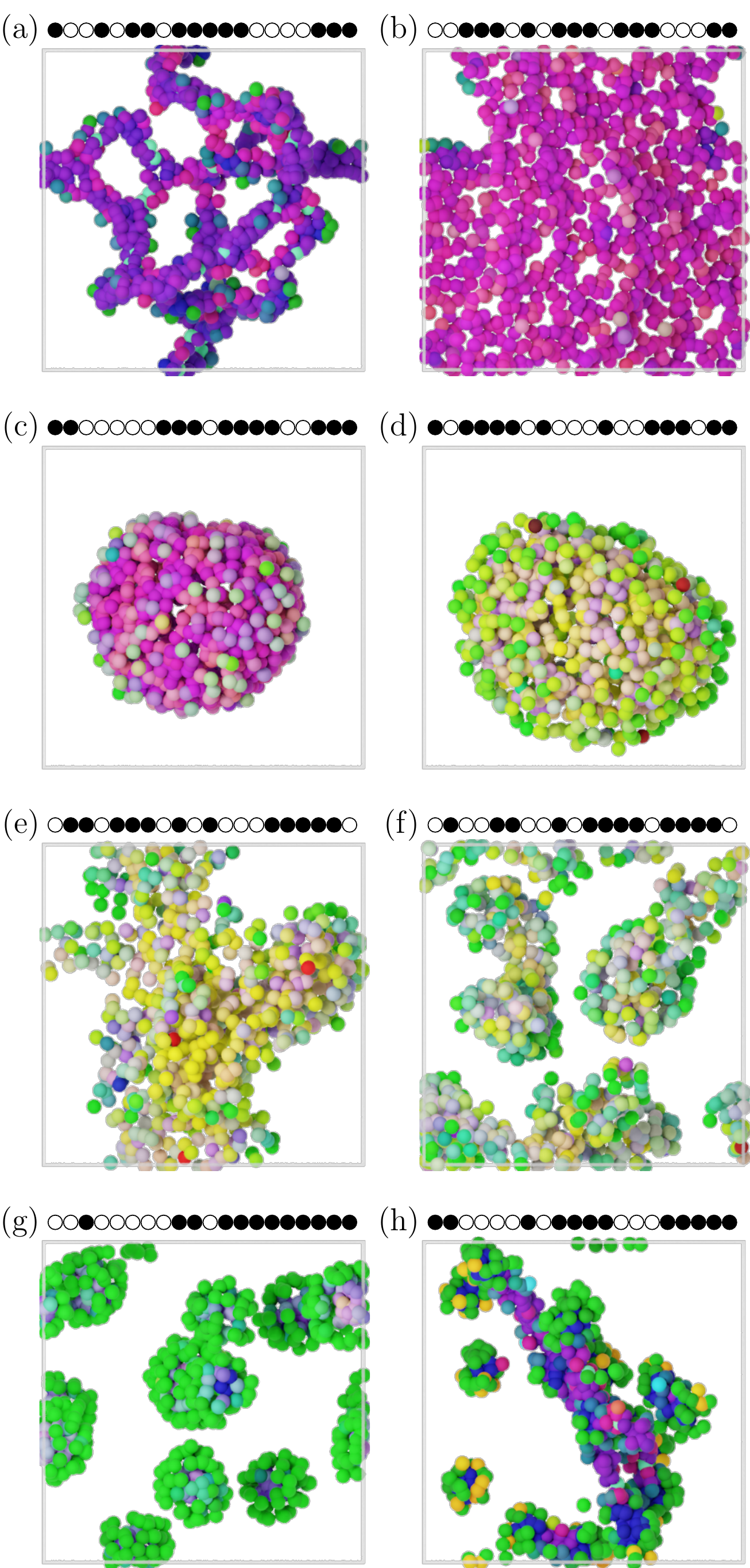}
    \caption{Representative structures observed from the manifold: Strings (a), membranes (b), vesicles (c), liquids (d), structured liquids (e), disordered micelles (f), spherical micelles (g), and wormlike micelles (h).
    Sequences shown schematically, with \A\, indicating an $A$-bead and \B\, a $B$-bead. Colors indicate positions of particle in $\mathbf{Z}^\circ$. Identification of structures was made based on the systems position in the $\mathbf{Z}$ latent space. 
    }
    \label{fig:structures}
\end{figure}

\begin{figure*}[t]
    \centering
    \includegraphics[width=\textwidth]{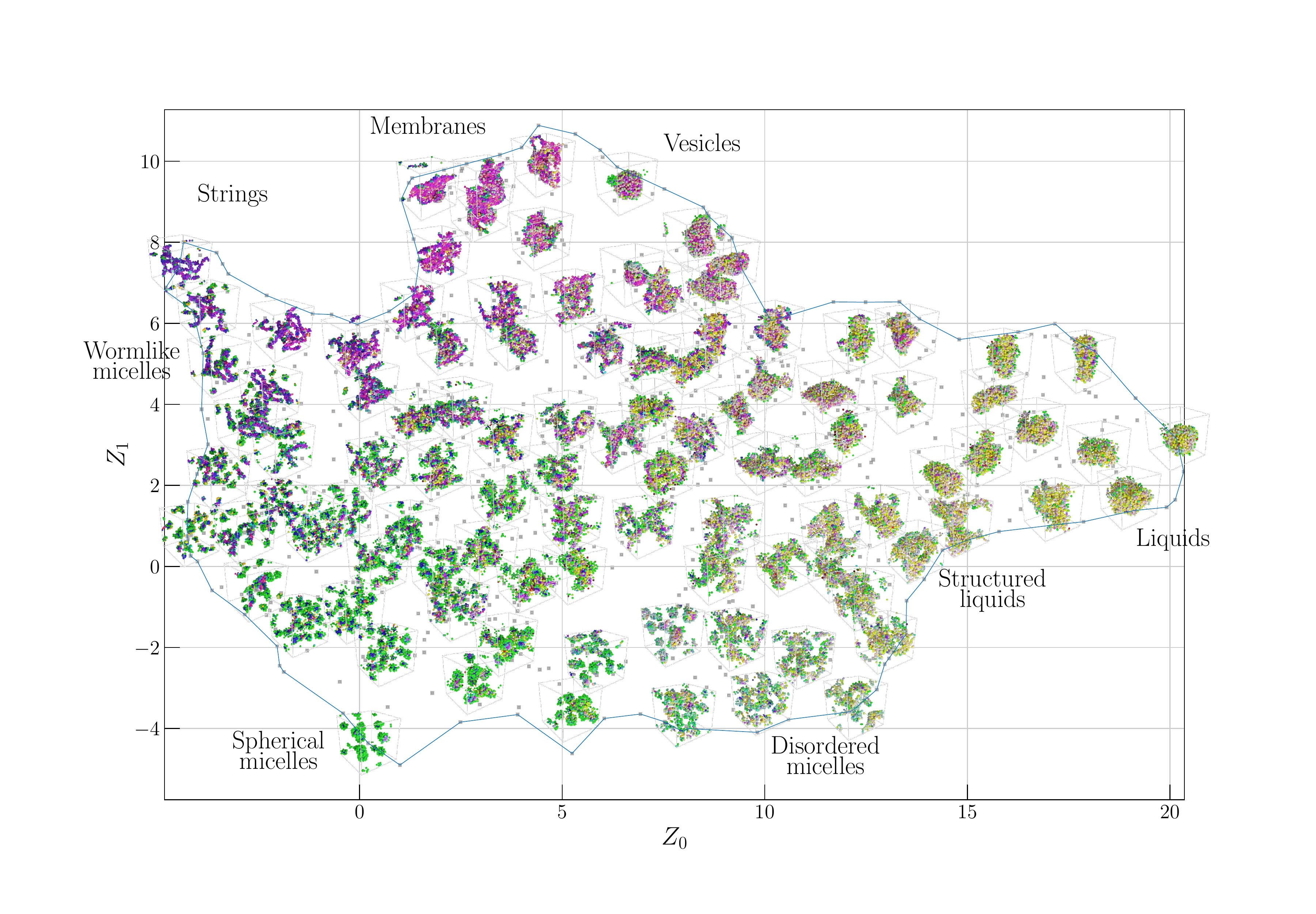}
    \caption{
    Manifold $\mathbf{Z}$ learned with UMAP on 1016 randomly selected structures.
    102 representative structures are pictured at their corresponding position in the manifold.
    The rest are indicated by symbols to avoid overlapping the renderings.
    Blue line indicates latent space periphery as determined by alpha-shape with $\alpha = 1.0$.
    }
    \label{fig:umap-random}
\end{figure*}

Even though the model investigated here is simple and only the sequence was varied -- even composition was fixed -- we found extremely diverse aggregate types.
Examples of these aggregate morphologies are shown in Fig.~\ref{fig:structures} (along with the corresponding monomer sequence), with each particle colored according to its characteristic local environment.
These colors correspond to the position in the particle latent space $\mathbf{Z}^\circ$ as shown in Fig.~\ref{fig:schematic} and in Fig.~S1 of the ESI\cite{SI}.
Purple color corresponds to a string-like environment, pink to sheet-like, yellow to liquid-like, green to micelle-like exterior, and blue to micelle-like interior.
These can be uniform throughout the simulation box (as in the top row) or highly heterogeneous (as in the bottom row), depending on the aggregate morphology.
The morphologies in Fig.~\ref{fig:structures} correspond well to prior observations in Ref.~\citenum{Statt2020}, which named most of the following aggregate types: strings (a), membranes (b), vesicles (c), liquids (d), structured liquids (e), disordered micelles (f), spherical micelles (g), and wormlike micelles (h).
Because of the way the morphologies were evaluated in Ref.~\citenum{Statt2020}, structured liquids and disordered micelles were grouped together.

To evaluate the range of possible aggregate morphologies, 1016 sequences were selected at random and equilibrated by MD for $2.5 \times 10^4 \, \tau$.
The embedding simulation system latent space $\mathbf{Z}$ for the resulting 1016 equilibrated snapshots is shown in Fig.~\ref{fig:umap-random}, along with renderings of some representative morphologies.
Because UMAP preserves the topology of the original high-dimensional data, a key feature is that the most dissimilar structures appear along the outer periphery of the manifold, while the interior samples interpolate between them.
This can be observed in the color distribution of each rendering, with one or two dominant colors around the periphery (e.g., all pink particles corresponding to all sheet-like environments), whereas in the center some snapshots contain a mix of many colors.

The `archetypical' morphologies then correspond to these peripherally located structures in $\mathbf{Z}$.
Starting at the top left and moving clockwise, these correspond to: strings, flat sheets, hollow vesicles, structured liquids, liquid droplets, disordered micelles, spherical micelles, worm-like micelles, then back to strings.
Towards the center of the manifold, mixtures of these archetypes appear, as both coexisting aggregates (e.g., part sheet / part string in the top right) and homogeneous mixtures (e.g., liquid-like disordered micelles in the bottom right).

\subsection{Sampling strategy\label{sec:sampling}}

\begin{figure*}[t]
    \centering
    \includegraphics[width=\textwidth]{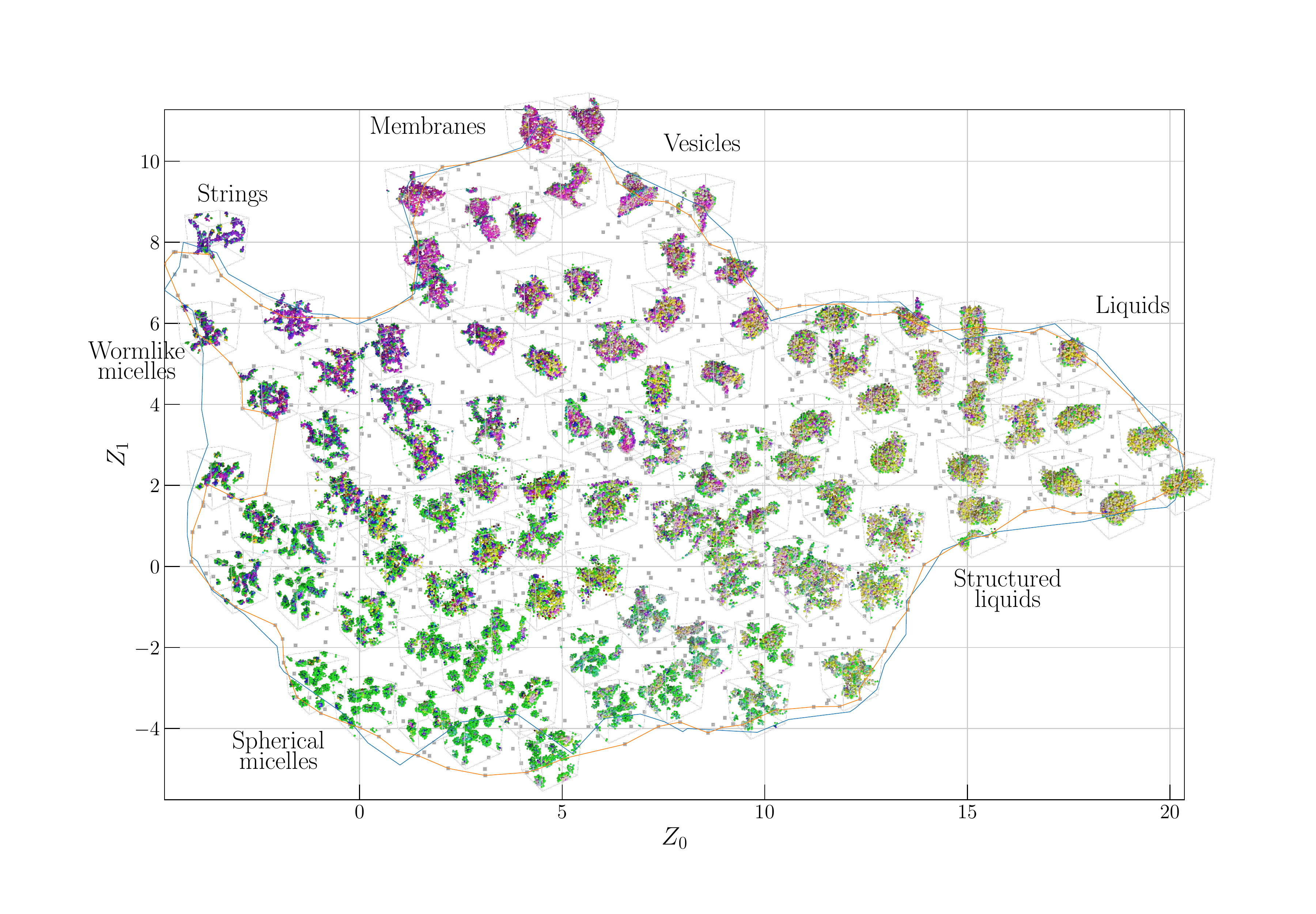}
    \caption{
    Manifold $\mathbf{Z}$  learned with UMAP on 1022 self-assembled structures.
    102 representative structures are pictured at their corresponding position in the manifold.
    The rest are indicated by symbols to avoid overlapping the renderings.
    Blue and orange lines indicates random and k-means latent space peripheries (respectively) as determined by alpha-shape with $\alpha = 1.0$.
    }
    \label{fig:umap-kmeans}
\end{figure*}

For such a diverse collection different morphologies, it is a challenge to ensure complete coverage of all aggregate types.
It is encouraging that the morphologies reported here already match those identified in Ref.~\citenum{Statt2020} using expert knowledge, such as selecting sequences with specific blocks and patterns or completely random ordering.
Nevertheless, given chains of length 20 with 12 $A$-type beads and 8 $B$-type beads, there are $63,090$ unique sequences to choose from, such that only $1.6 \%$ of possible sequences in this already highly constrained design space were evaluated to generate Fig.~\ref{fig:umap-random}.
Furthermore, there is no guarantee that expert knowledge and traditional measures like scattering functions are sufficient to identify and distinguish between all possible morphologies with so many possible sequences.

Here, we use two different sampling strategies to attempt good coverage of all aggregate types.
We have already shown the result when selecting sequences at random in Fig.~\ref{fig:umap-random}.
However, it is unlikely that the distribution of sequences has strong correspondence with the distribution of morphologies.
One obvious example is the relative rarity of blocky sequences -- there are 7 unique sequences with a block of 8 $B$-type beads, 78 with a maximum block size of 7, and 510 with a maximum block of size 6.
This continues to increase up to 25,731 for maximum block size of 2, but then drops down to only 651 sequences with no adjacent $B$-type beads.
Thus, sampling randomly offers more than an 80\% chance to select a sequence with a maximum block size of 2 or 3 $B$-type beads, such that both the very blocky sequences and the most non-blocky sequences are greatly under-represented.

To test the effect of sampling strategy, we first need to define a different distribution based on which the sequences are selected.
This estimate should be inexpensive since it must be computed for all $63,090$ sequences.
We therefore simulated clusters of 5 chains in short bursts of $50 \, \tau$ by MD.
The average LJ interaction energy $E_\text{LJ}$ and average radius of gyration $R_{g}$ of chains in these clusters were selected as proxies for the aggregation behavior.
K-means clustering was used to select 1022 sequences which were uniformly spaced over this $(E_\text{LJ}, R_g)$ feature space.
The resulting manifold is shown in Fig.~\ref{fig:umap-kmeans}.

There are two obvious distinctions between the randomly selected set in Fig.~\ref{fig:umap-random} and the k-means set in Fig.~\ref{fig:umap-kmeans}.
First, the density of spherical-micelle-forming sequences is much higher in the k-means set compared to the randomly selected set.
This is consistent with the above observation of the relative rarity of blocky sequences and the common knowledge that blocky sequences form micelles \cite{Hugouvieux2011,Floriano1999}.
Second, there is a collection of structures on the far left of the random manifold that is missing from the k-means manifold, with a notable concave region between the strings and wormlike micelles.
This implies that these morphologies are formed by sequences with short blocks of $B$-type beads, which are vastly over-represented when choosing sequences at random.

In the rest of this paper, we will focus our attention on this k-means set, as it appears to provide a more uniform distribution over the various morphologies.
However, the question of how to ensure coverage of the morphology space is far from answered, and future work with these sequence defined macromolecules should consider this problem carefully.

\subsection{Comparison to expert knowledge \label{sec:expert_knowledge}}

\begin{figure}[t]
    \centering
    \includegraphics[width=8.5cm]{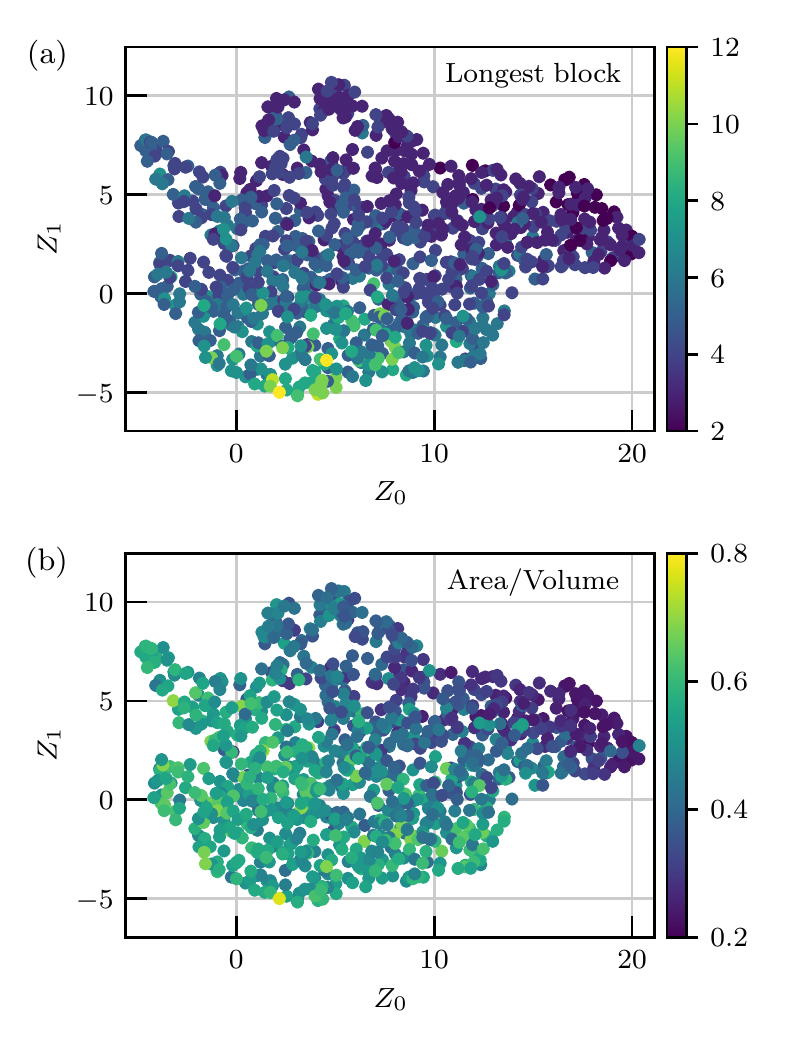}
    \caption{Manifold $\mathbf{Z}$  with each sequence colored by (a) the longest block of the same monomer and (b) the area-to-volume ratio used in Ref.~\citenum{Statt2020}.}
    \label{fig:umap-seq-features}
\end{figure}

Prior attempts~\cite{Statt2020} of distinguishing these aggregates morphologies looked at measures like pair-correlation functions (which can be related to the structure factor via Fourier-transform), number of nearest neighbours, and single-chain properties like $R_g$.
All those short-ranged structural measures showed very similar behavior; they usually distinguish between micelles (i.e., small clusters) and non-micelle structures (i.e., large aggregates), but are unable to identify subtler structural variations.

In an attempt to include global structure, we~\cite{Statt2020} have previously evaluated these morphologies by a surface mesh method \cite{Stukowski2014}.
This results in a surface defined by a triangulated mesh enclosing the dense aggregates.
The area to volume ratio ($A/V$) distinguished the liquids, characterized by large aggregates with low $A/V$, from the spherical micelles, characterized by many smaller aggregates with corresponding high $A/V$.
The genus $G$ -- a topological measure of a surface -- was also used to distinguish structured liquids (negative $G$, corresponding to internal voids) from strings (positive $G$, corresponding to holes).
The area-to-volume ratio and genus (measuring the topological properties of the mesh) was then used to classify morphologies. Note that this way of identifying aggregate types lead to disordered micelles and structured liquids being grouped together, since both have a significant number of voids and similar $A/V$ ratio. 

These measures were able to roughly characterize the global structure of each archetype, but they do not account for mixtures of the structures as observed in the center of Figs.~\ref{fig:umap-random}-\ref{fig:umap-kmeans}.
This is due to the global pooling of the mesh measures, which reduces the entire feature space to only two numbers.
In the current approach, we retain information about the distribution of individual particle environments over the entire aggregate and only reduce to two-dimensional latent space at the end.

Interestingly, the archetype morphologies appear to be determined by the features of the local environments.
This is especially true for strings and membranes, which appear in solid purple and pink, respectively, with only a few green particles around the edges.
Spherical micelles have two distinct environments corresponding to the core and shell regions, but the local environments fall cleanly into one of these two.
This suggests that the global morphology is a direct consequence of the local aggregation behavior of the chains.
However, we also reiterate that the characterization presented in Figs.~\ref{fig:umap-random}-\ref{fig:umap-kmeans} is not the result of analyzing local structure alone, but also the distribution of local structure observed over all chains in the system.

To assess the ability of the sequence characteristics to predict the self-assembled morphology, we evaluated some sequence order parameters in line with Ref.~\citenum{Statt2020}.
The most salient feature was the longest block of $A$-type beads $L_A$ (equivalent to $M - L_B$), which is plotted in Fig.~\ref{fig:umap-seq-features}.
Here, spherical micelles are clearly distinguished by their very long blocks -- these correspond to only a few blocks, either di-block or tri-block.
Other structures are less clear, with only one dimension of the manifold being reliably provided by this measure -- it seems that short blocks correspond to either liquids or membranes/vesicles, but distinguishing between those is challenging.

When comparing how other short-ranged structural properties are distributed in the manifold, they all show similar behavior (see Fig.~S4 in the ESI\cite{SI} for additional measures).
A diagonal gradient in latent space is observed, reflecting the ability to distinguish many small clusters from few large ones.
This also roughly corresponds to the $A/V$ ratio of the final structures as shown in Fig.~\ref{fig:umap-seq-features}.
The cluster of similar green color in the bottom left half of the manifold demonstrates the degeneracy of this metric when dealing with more nuanced structures.
Thus, the features produced by unsupervised learning provide additional information beyond what expert knowledge provides.

\subsection{Finite size effects \label{sec:finite-size}}

\begin{figure}[!t]
    \centering
    \includegraphics[width=8cm]{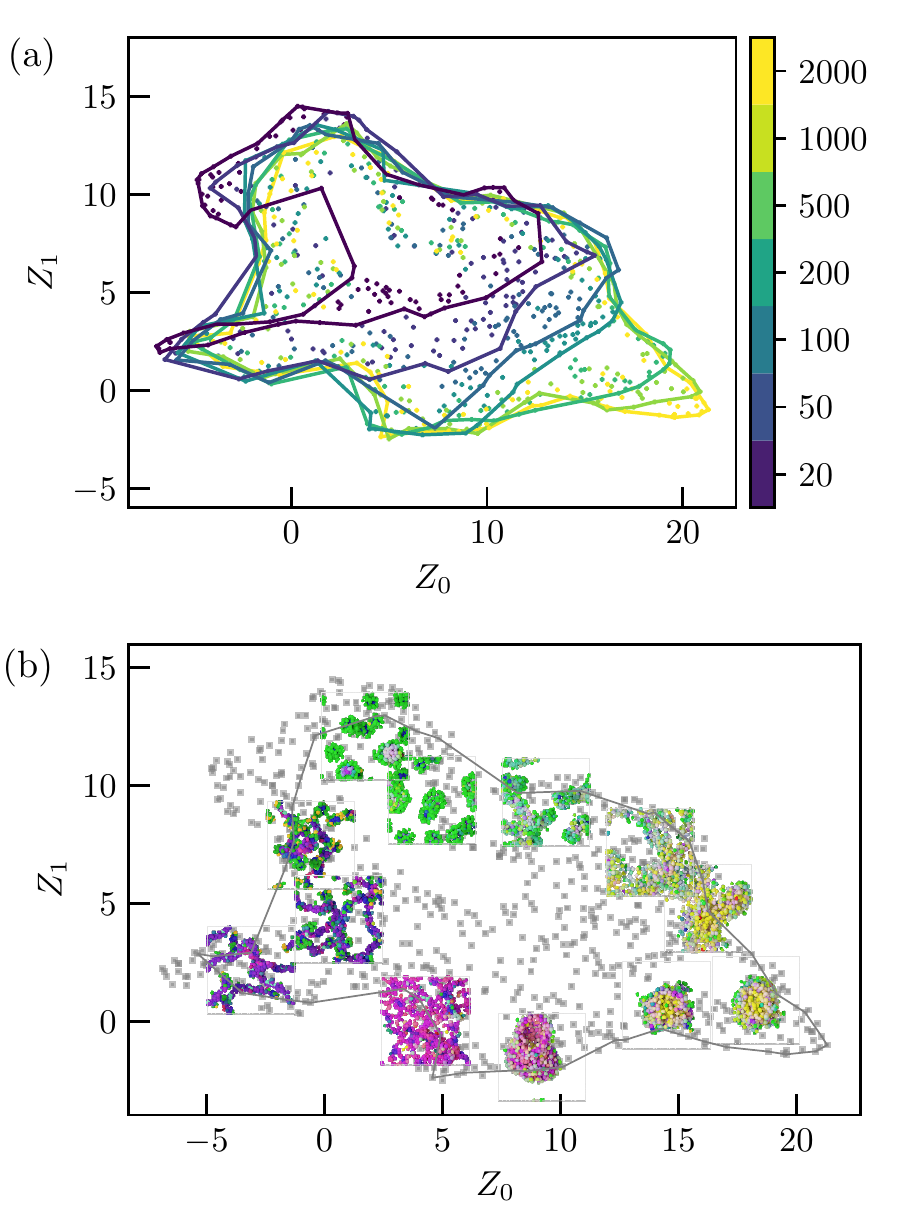}
    \caption{
    (a) Point clouds and enclosing alpha-shapes ($\alpha = 0.4$) for peripheral sequences with 20, 50, 100, 200, 500, 1000, and 2000 chains fit to a new manifold (all at once).
    (b) Representative snapshots embedded in the $\mathbf{Z}$ manifold from only $N = 500$ chains.
    }
    \label{fig:concentric}
\end{figure}

As always with molecular simulation, it is critical to evaluate the effect of different system sizes on the final results.
This is especially true when considering large-scale aggregates, which may require a very large number of chains to realize the full range of possible structures.
Yet for computational efficiency, a smaller number of chains is preferable, especially when considering the thousands of different sequences that were evaluated here.
We also have some evidence from the preceding sections that the global structure is informed by local environments (i.e., that archetypes are comprised mostly of the same color particles).
This leads to the question: At which system size does the full range of possible morphology archetypes appear?

To answer this question, we picked representative sequences from the periphery of the manifold and repeated simulations with the following system sizes: 20, 50, 100, 200, 500, 1000, and 2000 chains.
All systems were evaluated in a cubic box at a fixed density of $0.05\,\sigma^{-3}$.
Sequences from the periphery were chosen because they represent the archetypical morphologies and should always define the outer boundaries of the manifold; this hypothesis was validated on the $N = 100$ system in the ESI\cite{SI}.
This strategy allows a measure of the coverage in the morphology space while greatly reducing the computational burden of simulating all sequences at all system sizes.

From the simulations at each system size, a new manifold was constructed using all results at once.
This is to avoid biasing the result based on a reference system size, since manifold learning performs substantially better at interpolation than extrapolation.
The resulting manifold is shown in Fig.~\ref{fig:concentric}, with points labeled for each system size in panel (a) and examples rendered for the $N = 500$ system size in panel (b).

Note that the resulting manifold is flipped along the $Z_1$ axis relative to the manifolds shown above in Figs.~\ref{fig:umap-random}-\ref{fig:umap-kmeans}, since the latent space mapping is not unique.
This is a common feature of manifold learning approaches, which are tasked with preserving the topology of the high-dimensional data but rarely are provided objectives which prioritize certain orientations.
Thus, the nearly constant topology of these mappings is meaningful whereas the sign flip is not.
Comparing manifolds generated from different subsets of the data can be achieved by a Procrustes analysis or by embedding the data from one manifold into the other.
For classification workflows, this does mean that a few labels are required to ascertain a suitable alignment; these could even be the same few snapshots which are always embedded into each manifold as ``landmarks.''
In that case, the proposed method would be considered semi-supervised.

The alpha-shapes for each system size shown in Fig.~\ref{fig:concentric}(a) reveals a gradual increase of latent space coverage with increasing system size.
For 20 and 50 chains, most sequences result in small aggregates which lack clear definition as one of the archetypes.
It especially appears that string, membrane/vesicle, and liquid-like morphologies are not able to form.
This may be an artifact of the coarse-graining procedure since these systems have only 40 and 100 total beads, respectively.
However, it is more likely evidence for a minimum system size required to achieve the full range of hierarchical structures accessible with more chains.
With 100 chains, most structures can form, but the variety of liquids and vesicles are restricted compared to larger systems (compressed boundary on the bottom right corner).
By 200 chains, all structures except some of the liquids are accessible, and above 500 chains little difference is observed in the coverage except a slightly larger liquid lobe.
This makes sense as larger systems can adopt more varied amorphous phases, whereas the homogeneous archetypical structures just have more of the same.

\subsection{Variability \label{sec:variability}}

In the preceding analysis, all simulations were initialized randomly, following the same randomly placed freely rotating chains procedure as Ref.~\citenum{Statt2020}.
Beyond randomness in the initialization, the self-assembly process is also stochastic at finite temperature and so the final structure formed by the same sequence may be different in different runs.
We therefore probed the variability in latent space in two ways: first, by running the same sequence from the same initial (random) configuration with different thermostat seeds, and second by initializing the same sequence in random different configurations.

\begin{figure}[!t]
    \centering
    \includegraphics[width=8cm]{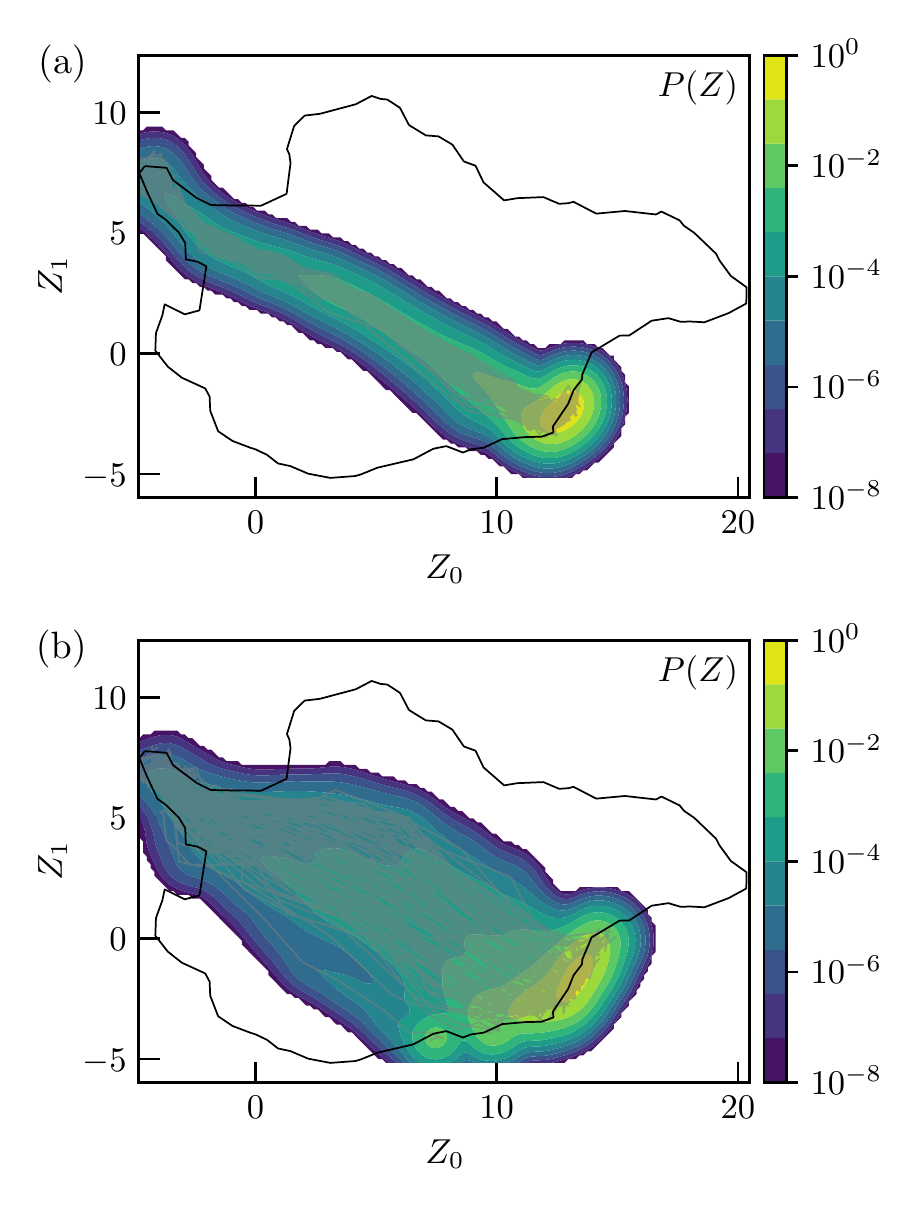}
    \caption{Trajectory through latent space $\mathbf{Z}$  for a spherical-micelle-forming sequence (see Table~\ref{table:sequences}).
    Probability density from 125 trajectories is shown with contours, while each individual trajectory is overlaid with lines.
    (a) initialized from the same configuration but using a different random seed for the thermostat and (b) initialized randomly.
    }
    \label{fig:replica}
\end{figure}

To evaluate the variability induced by random thermal noise (i.e., thermostat pseudo-random number seed), we ran 125 replicates for $5 \times 10^3 \, \tau$ and recorded their position in the latent space at 21 logarithmically spaced time intervals.
We observed that random initialization closely resembled strings, since the chains are distributed throughout the simulation box at low density.
The sequence for this test was therefore chosen from a previous trajectory which terminated in the bottom right of the latent space, such that equilibration is expected to drive the morphology across the latent space from top left to bottom right over time.
The trajectories are shown in Fig.~\ref{fig:replica}(a) by grey lines, with probability densities from kernel density estimation shown in the contours (with logarithmic levels).
We indeed saw trajectories moving between the expected two regions of latent space.
As expected, the thermostat seed induced little variability in either the path or the final position of the aggregates in latent space.

In the manifolds of Figs.~\ref{fig:umap-random}-\ref{fig:umap-kmeans}, final aggregates were also subjected to random placement of the chains between each sequence.
Therefore, it is critical to evaluate the variability induced by this random initialization given the same sequence.
Note that this influence is effectively stacked with the random thermostat seed since it is physically meaningless to have the same thermal noise with a different configuration.
As before, the trajectories are shown in Fig.~\ref{fig:replica}(b) by grey lines, with probability densities from kernel density estimation shown in the contours.
Here the variability is substantially greater between replicas, with a broad range of intermediate states in the center of the manifold leading to an equally broad final state.

Reassuringly, these final positions all correspond to the same morphological archetype, a disordered micelle positioned in between spherical micelle and liquid droplet.
This result also invites a discussion about the assumptions made during UMAP embedding, namely that UMAP assumes uniform density in the manifold.
Here we think the disordered states offer substantially greater variety in the possible states which roughly correspond to the same archetype compared to strings, spherical micelles, or membranes.
This has already been addressed to some extent in Section~\ref{sec:finite-size}, and will be further evaluated further by in the following Section~\ref{sec:trajectories}.

\subsection{Dynamics in latent space \label{sec:trajectories}}

In a complementary manner to the variability from random configurations at a single point in latent space, we can also consider how the equilibration of a particular sequence is affected by initialization at different points in latent space.
Here we use 125 configurations defined by the alpha shape of the k-means manifold in Fig.~\ref{fig:umap-kmeans} as starting states, but switch to a new polymer sequence and then relax with MD again.
This creates a trajectory through latent space from initial starting points from all around the edges of the manifold towards a single equilibrium point (or region).
From these trajectories, we obtain information about the variability exhibited by a single sequence, comparable to Fig.~\ref{fig:replica}(b), but also evaluate the smoothness of the real MD trajectories through the latent space.

\begin{figure}
    \centering
    \includegraphics[width=85mm]{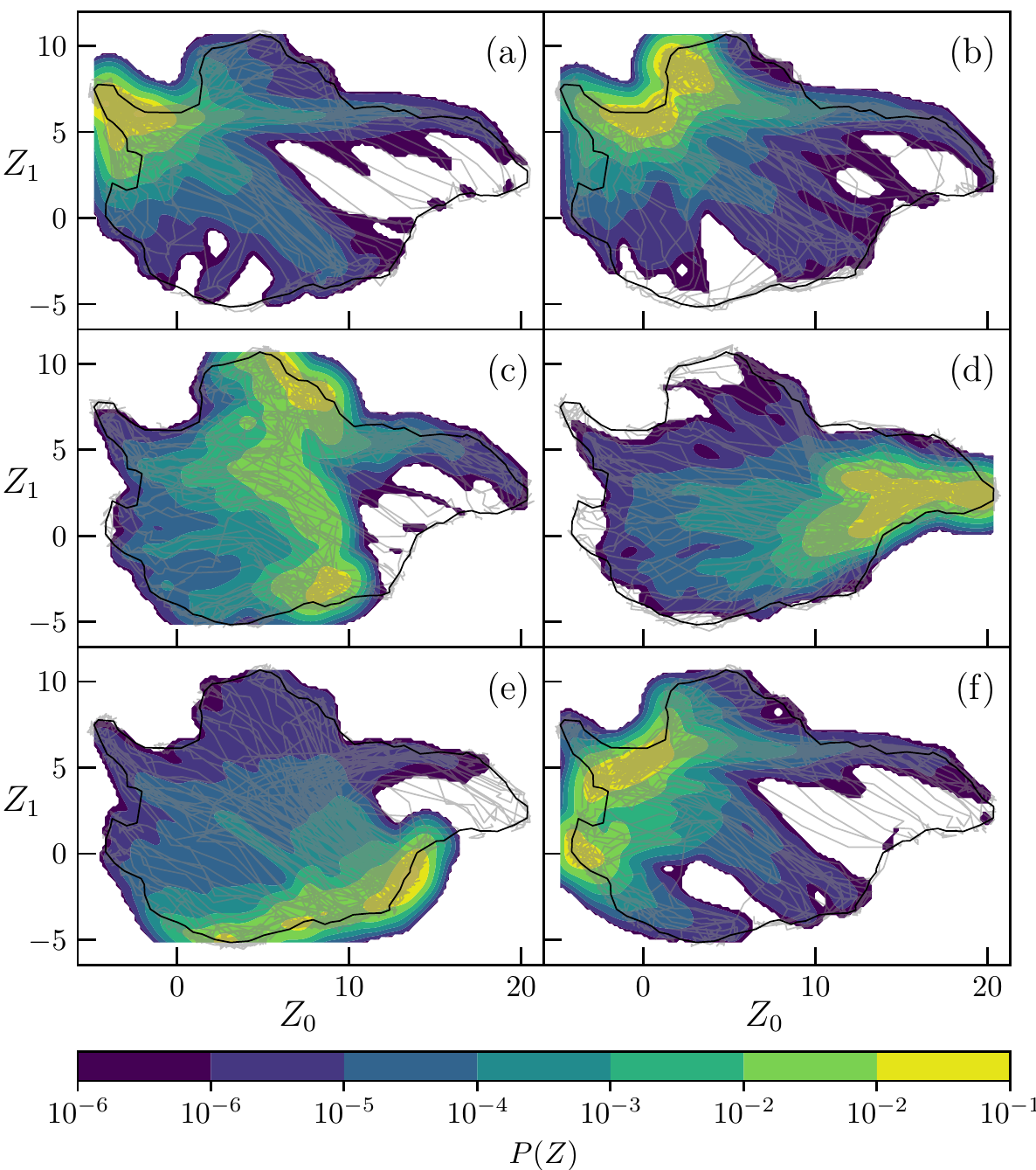}
    \caption{Probability distribution in latent space for various sequences initialized as other structures.
    (a) string, (b) membrane, (c) vesicle, (d) liquid, (e) spherical micelle, (f) wormlike micelle.
    The sequences are listed in Table~\ref{table:sequences}.
    }
    \label{fig:trajectory}
\end{figure}

The trajectories from $5 \times 10^3 \, \tau$ are shown in Fig.~\ref{fig:trajectory}, with a similar visual scheme as Fig.~\ref{fig:replica}.
Note that the trajectories had to be sampled with logarithmic spacing in time in order to obtain smooth curves because their velocity through latent space is very high far from their equilibrium positions, but slows down significantly over time.
There are clearly preferred pathways between different regions of the latent space, but hysteresis again leads to variations in the path and final morphology.

Most of the sequences end up in a fairly tight region of latent space, providing evidence that this equilibration procedure is reliable and kinetic trapping is fairly rare.
Strings collapse to a highly localized distribution, whereas liquids and spherical micelles are broad -- this supports the idea that the structural latent space may not really have a uniform density.
In the case of membranes and wormlike micelles there are two distinct minima but they are directly adjacent.
However, for vesicles, a significant hysteresis is observed between true vesicles and small, spherical-micelle-like aggregates.
While some consistency is observed in the path of the initially liquid-like states, the initially wormlike micelle states appear to randomly diverge up to vesicles or down to spherical micelles.

\begin{figure}
    \centering
    \includegraphics[width=85mm]{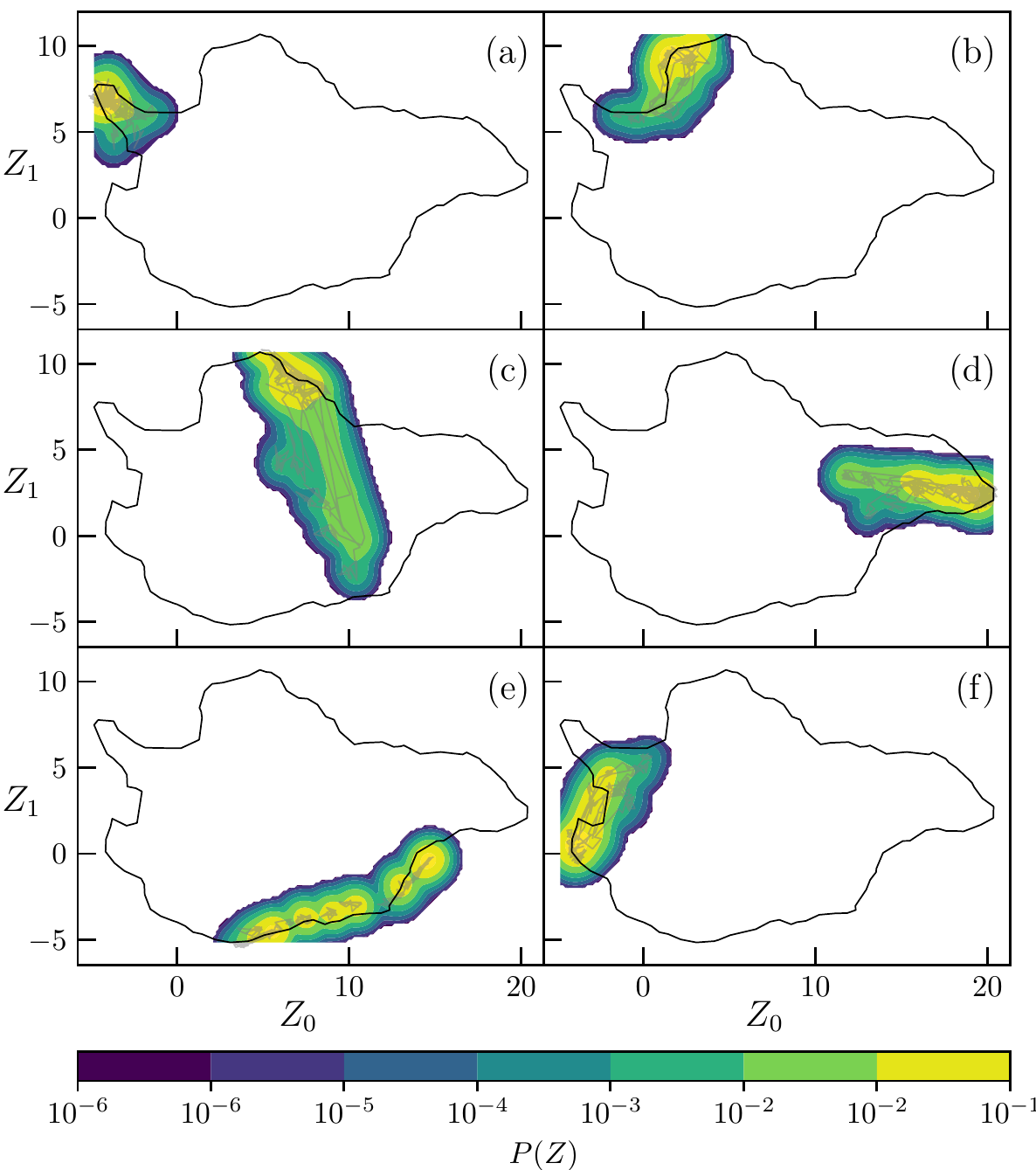}
    \caption{Probability distribution in latent space for various sequences initialized from those in Fig.~\ref{fig:trajectory}
    (a) string, (b) membrane, (c) vesicle, (d) liquid, (e) spherical micelle, (f) wormlike micelle.
    The sequences are listed in Table~\ref{table:sequences}.
    }
    \label{fig:trajectory-extended}
\end{figure}

To provide further insight into this phenomenon, a subset of the simulations from above were continued for an additional $2 \times 10^5 \, \tau$.
This was intended to evaluate the metastability of the observed states; we do not really know if these morphologies are equilibrium or kinetically trapped structures.
Interestingly, the additional equilibration time showed clear convergence to a single minimum for all except the spherical-micelle-forming sequence.
Each vesicle-forming sequence eventually escaped the spherical-micelle-like minimum to form a vesicle morphology.
Likewise, the dual minima of the membranes and wormlike micelles converged somewhere in between.
Even the liquid states became more ``liquid-like'' (moving to the far right of the manifold).
However, in the case of the spherical-micelle-forming sequence, the aggregates stayed confined to their initial regions, indicating that they are metastable.

A few key conclusions can be drawn from these trajectories for this paticular system.
Perhaps most importantly, the trajectories through latent space are smooth with respect to the simulation time, such that a continuous time evolution by MD corresponds to a smooth evolution through the collective variable $Z$.
This was not necessarily the case by construction, since the manifold was fitted from simulations of different sequences and included only one observation from each such simulation. 
Conversely, the space (distinguishing different large scale aggregates) appears smooth between sequences, with different trajectories initialized near to each other following similar paths; there do not appear to be discontinuities in the space. 
This also indicates that the determined large-scale aggregate structures may not strictly belong into well separated categories (like different crystal structures would), but span a more continuous, gradual space.  
hile the shown probability distribution in latent space are not free energy surfaces, the fact that the latent space appears to be spatiotemporally smooth may offer interesting ways of constructing free energy surfaces in the future. 

\begin{table}[hpt!]
\caption{ Sequences used for latent space trajectories, where \A\, indicates an $A$-bead, and \B\, a $B$-bead. The longest block $L_A$ of A beads as well as the number of A-blocks $n_A$ is given.
\label{table:sequences}
}
\begin{tabular}{ c c c c c }
 Morphology & Sequence  & $L_A$ &  $n_A$ \\
 \hline
 string  & \A\B\B\A\B\A\B\A\A\A\A\A\B\A\B\B\A\B\A\A  & 5 & 7 \\ 
 membrane  & \A\B\A\A\A\B\B\A\A\A\A\B\B\A\B\B\B\A\A\A & 4 & 5\\
 vesicle  &\B\A\B\B\B\A\A\B\A\A\A\A\B\A\A\B\A\A\B\A & 4 & 6\\
 liquid  & \B\A\A\B\A\A\B\A\A\B\A\A\B\A\B\B\A\A\A\B & 3 & 6 \\
 spherical micelle  & \B\B\B\B\B\B\B\B\A\A\A\A\A\A\A\A\A\A\A\A & 12 & 1 \\
 wormlike micelle  & \B\B\A\A\A\B\B\B\A\A\A\A\B\B\B\A\A\A\A\A & 5 & 3 \\
 \hline
 \end{tabular}
 \end{table}

\section{Conclusions\label{sec:conclusions}}

In this work, we demonstrate the application of unsupervised machine learning of local environments with simple rotation-invariant features, as recently described in detail in Ref.~\citenum{Reinhart2021}, to large-scale aggregates of a model copolymer.
Prior unsuccessful attempts~\cite{Statt2020} at characterizing these aggregates have relied on hand-crafted local structural order parameters.
The best previous approach classified the morphology of the large scale aggregates based on global information such as area and volume of a surface mesh.~\cite{Statt2020}
It is notable that the present method is able to characterize global structure in disordered aggregates based on information from the local environments, without explicitly considering global geometry (i.e., a mesh). 

The histogram-based pooling scheme of this method provides more nuanced descriptors than the mesh-based approach since local features (e.g. differently colored environments from $\mathbf{Z^\circ}$ in Fig.~\ref{fig:structures}) are considered as a distribution rather than as scalar quantities.
While in this work we focused on small systems, this feature would be especially advantageous for very large systems where blends of different aggregate types might be present and averaging would lead to significant information loss.

By evaluating over $2 \, 000$ different sequences, we provided additional evidence that the archetypes identified in Ref.~\citenum{Statt2020} are a nearly complete set.
We also showed that the sampling method for sequence selection can influence the structure of the resulting manifold based on the assumption of constant density, as highlighted in Fig.~\ref{fig:umap-kmeans}.
In contrast to prior expert classification, we were able to distinguish a disordered micelle aggregate type (see Fig.~\ref{fig:structures}(f)), which lies in between structured liquids and spherical micelles.
Based on our results, including a thorough finite size analysis, we believe that we have discovered all of the aggregate archetypes which can be assembled for the particular parameters used in Ref.~\citenum{Statt2020} ($f_A=0.6$, $M=20$, $T=0.5$).

Beyond confirming prior results, we showed that UMAP has several useful features in the context of identifying aggregate morphologies.
First, `archetypical' structures appear on the periphery of the manifold while mixed structures appear in the middle, which is a necessary consequence of the topological structuring enforced by the UMAP algorithm.
Second, the latent space learned from snapshots of different sequences is also continuous in time during a single trajectory.
This makes the UMAP latent space useful as a collective variable, and could lead to a useful parameterization of a free energy surface.

By investigating the evolution of a system from a particular starting point to its preferred equilibrium structure in the latent space, we showed that aside from the formation of spherical micelles from vesicle-forming sequences, no significant kinetic trapping occurred when evolving between structures.
We also observed that the structural evolution during very short times (less than $10^3 \, \tau$) can predict long-time aggregate structure (greater than $10^5 \, \tau$).
This is consistent with our observation that structural signatures from the local environments largely dictate the global structure (at least for the archetypical morphologies).

In this study, we have investigated only monodisperse chains with a single composition in order to evaluate the effect of sequence alone.
In experiments, polydispersity in chain length and variance in monomer sequence is expected.
Incorporating these effects into the analysis are promising next steps.
In addition, polydispersity might be able to be utilized to drive certain self-assembly pathways and might be exploited as a design parameter.
We also expect that the targeted inverse design of large-scale disordered aggregates using machine learning approaches might have interesting applications, particularly for drug delivery~\cite{Dutta2020,Posocco2010}. 

\begin{acknowledgments}
This work was supported in part by the Institute for Computational and Data Sciences and the Materials Research Institute at the Pennsylvania State University.
\end{acknowledgments}

\appendix
\section{Local environment UMAP}

\begin{figure}[h]
    \centering
    \includegraphics[width=\textwidth]{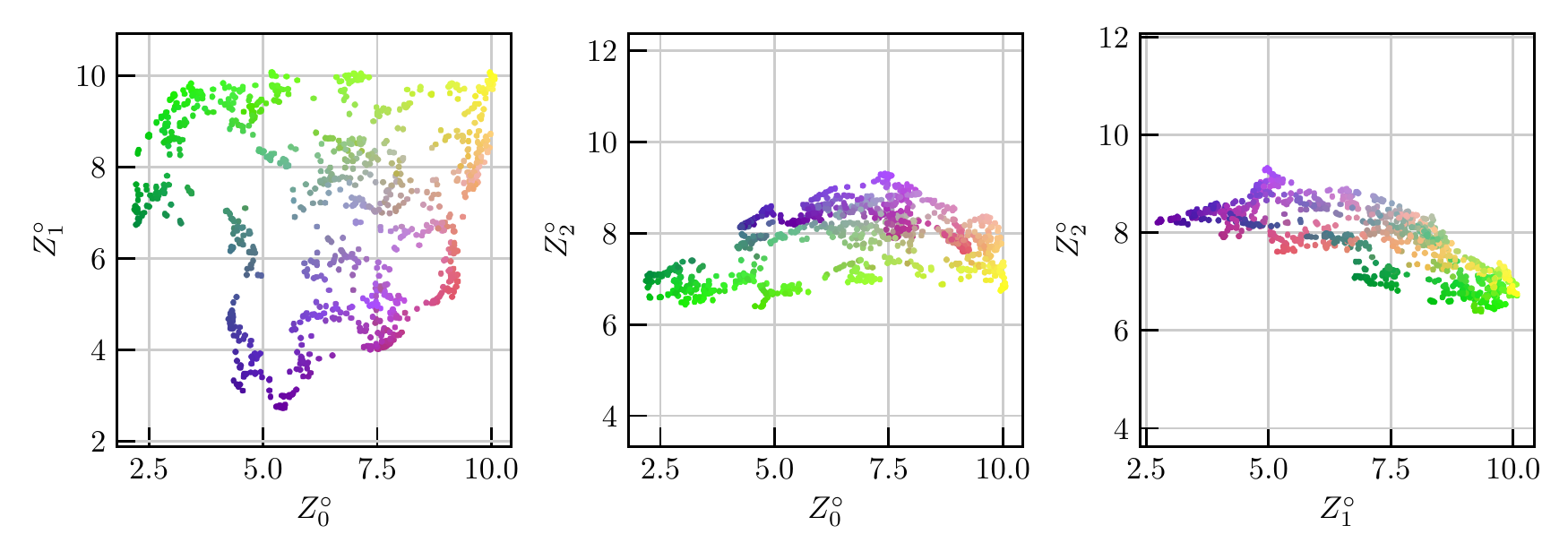}
    \caption{Manifold obtained by UMAP for local (coarse-grained) environments, with HSV color scale for easy visualization in the simulation snapshots.
    \label{fig:local-embedding}}
\end{figure}

Fig.~\ref{fig:local-embedding} shows the UMAP result for all local environments observed from the 1022 k-means sequences.
There were 1000 environments from each sequence (2 per chain), taken from a single snapshot at the end of the equilibration by MD.
The embedding was fixed for all other calculations of $\mathbf{H}^\circ$.
A color scale was assigned based on a Hue-Saturation-Value mapping from the 3D coordinates in $\mathbf{Z}^\circ$.
This color scale was applied to all particles shown in the main text, with detailed snapshots available in Fig.~2.

\section{Finite size effects}

\begin{figure}[h]
    \centering
    \includegraphics[width=0.67\textwidth]{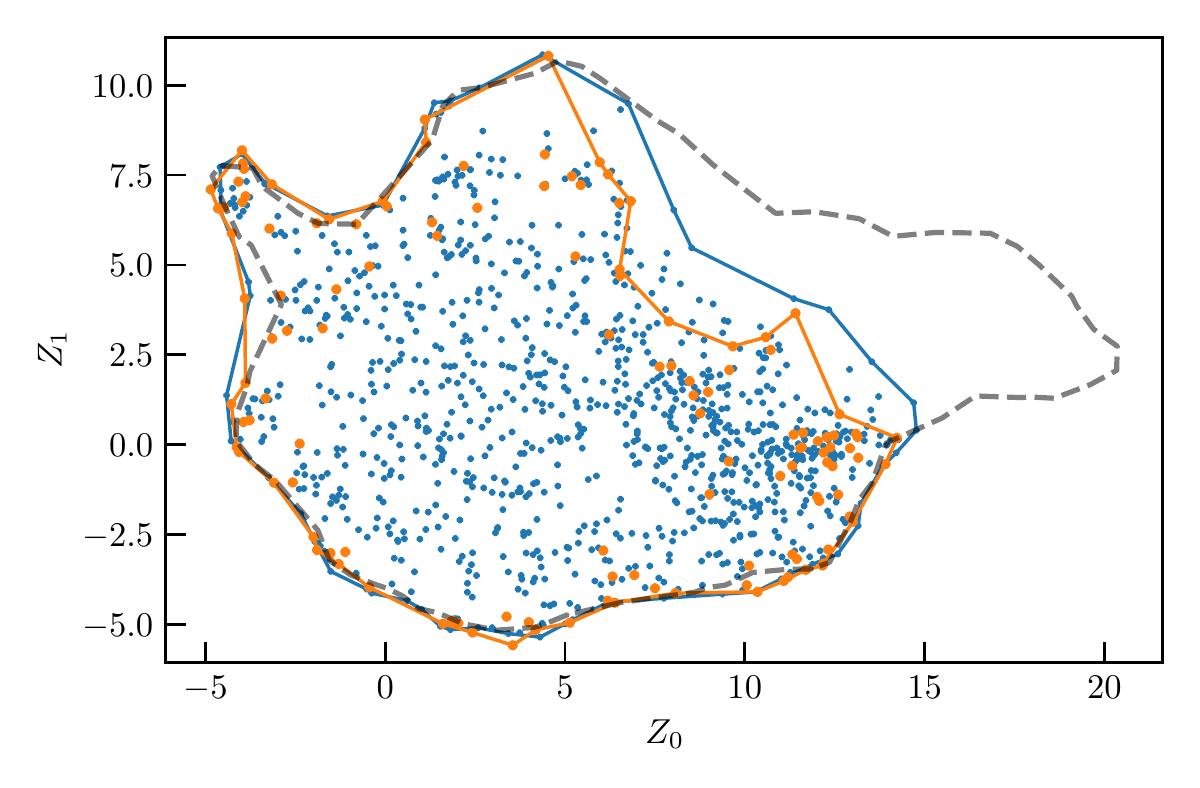}
    \caption{Manifolds with $N=100$ chains embedded into the k-means $N=500$ manifold.
    Blue symbols are all 1022 $N=100$ sequences, with blue line indicating the alpha-shape.
    Orange symbols are 125 peripheral $N=100$ sequences, with orange line indicating the alpha-shape.
    Dashed black line corresponds to alpha-shape for all data at $N=500$.
    }
    \label{fig:peripheral-100}
\end{figure}

In Fig.~\ref{fig:peripheral-100}, we show morphologies obtained from 100 chains for all 1022 sequences determined by k-means.
The 125 peripheral sequences are highlighted to show that these remain close to the periphery and are a reliable way to obtain the alpha-shape.
This provides a $8 \times$ time savings compared to simulating all 1022 sequences.
It is preferable to include only these peripheral sequences because they represent the morphological archetypes and most of the sequences in the center of the manifold are mixtures of those on the periphery.

\begin{figure}[h]
    \centering
    \includegraphics[width=0.50\textwidth]{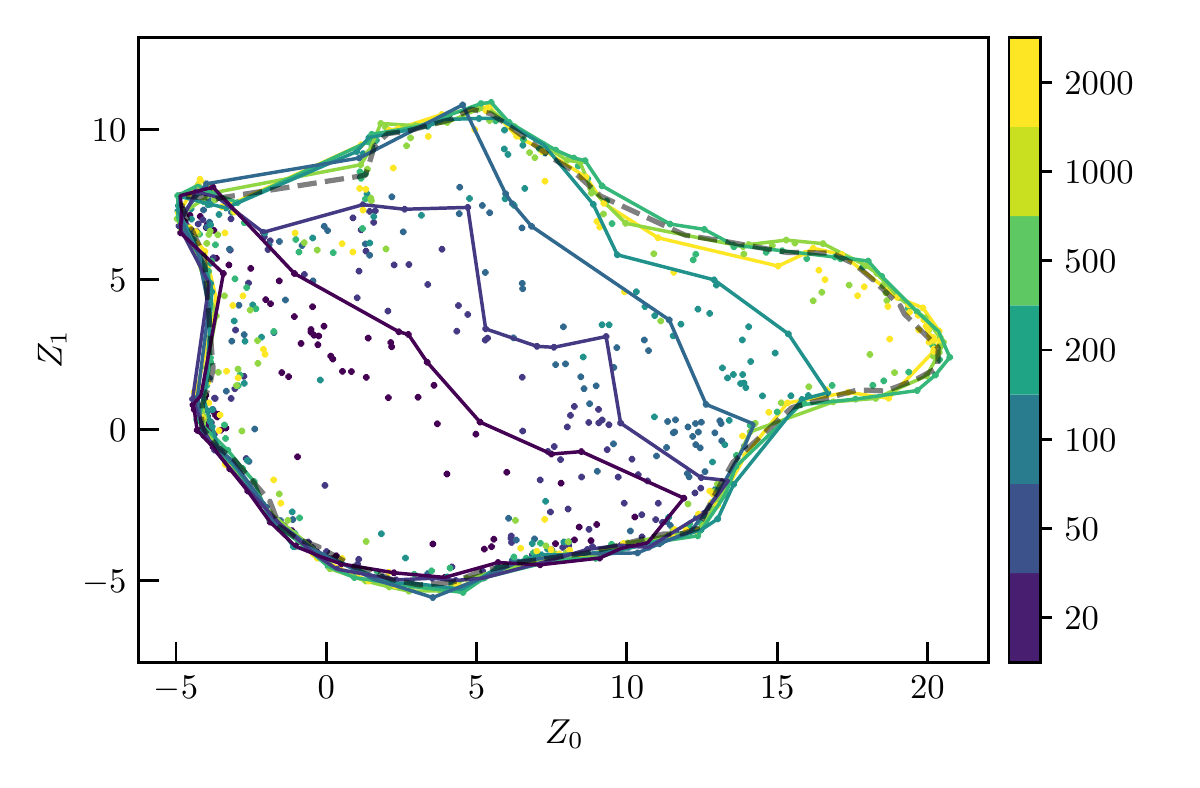}
    \caption{Morphologies for $N=$ 20, 50, 100, 200, 1000, and 2000 chains embedded in the k-means manifold from $N=500$ chains.
    Symbols correspond to each observed periphery sequence.
    Lines correspond to alpha-shapes for each $N$ (peripheral sequences only), with dashed black lines corresponding to alpha-shape for all data at $N=500$.
    }
    \label{fig:finite-size}
\end{figure}

In the main text, a new manifold was computed for the finite size study using only the peripheral sequences.
This was important because UMAP does not provide robust placement of observations outside of the training domain.
In essence, extrapolated points have no guarantee of topology preservation when projected into the learned low-dimensional space.
Nevertheless, for ease of comparison we show the morphologies from the finite size study embedded into the $N = 500$ manifold here.
As observed in Fig.~6, we see here that for low $N$, membranes, vesicles, and liquid droplets cannot form.
For $N=20$, wormlike micelles also cannot form which is visible from the location of the purple points, although the alpha-shape does not capture the inward deflection at the left side.
As $N$ increases, the right edge of the alpha-shape moves up and to the right (towards vesicles and liquids), and eventually the full space is recovered.
Note that unlike in Fig.~6, complete convergence appears to occur for $N=500$, demonstrating the extrapolation problem described above.

\section{Details from the embedding procedure}

The main text gave a brief description of the acquisition of \textit{features} and the procedure for \textit{pooling} into \textit{histograms} since this procedure follows closely the one introduced in our recent prior work \cite{Reinhart2021}.
Here we provide more details for representative examples to illustrate the procedure more explicitly for the macromolecular aggregates use case.

\begin{figure}[h!]
    \centering
    \includegraphics[trim=0 14cm 0 0, clip, width=0.48\textwidth]{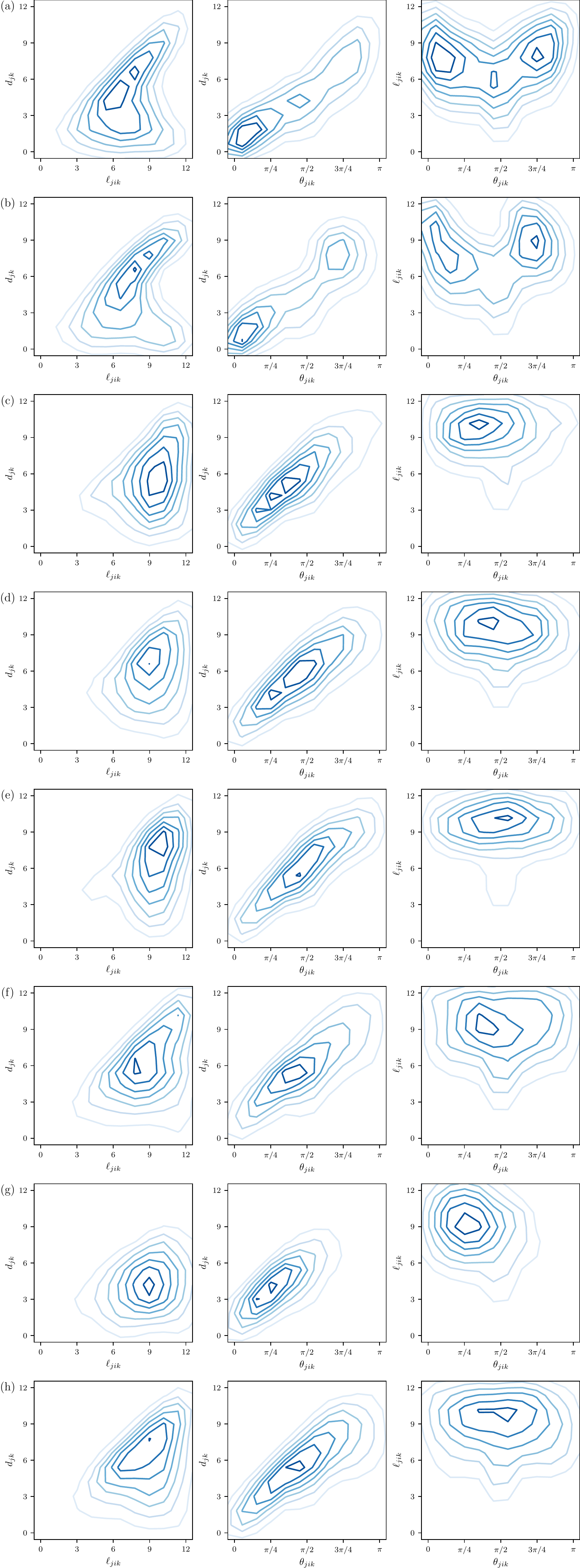}
    \includegraphics[trim=0 0 0 14cm, clip,width=0.48\textwidth]{figures/key-structures-features.pdf}
    \caption{Sample feature histograms (i.e., $\mathbf{H}^\circ$) corresponding to the structures in Fig.~2 in the main text.
    From each snapshot, a representative particle with the most common features (as determined by a kernel density estimation on the particle-level manifold) was selected.
    The three sub-panels for each structure represent different slices of the 3D histogram.
    Readers should refer to Ref.~\cite{Reinhart2021} for additional details and examples.
    }
    \label{fig:key-features}
\end{figure}

The histograms for each particle environment are shown in Fig.~\ref{fig:key-features}.
Each set of three sub-panels corresponds to a single representative local environment.
The raw features $\mathbf{F}$ are just an array and would be difficult to visualize, so this is after the first pooling step in Fig.~1.
These histograms will be flattened into a single feature vector for embedding in the next step.
The embedding occurs by comparing each of these feature vectors to each other and constructing a 3D vector in $\mathbf{Z}^\circ$ space, as shown in Fig.~\ref{fig:local-embedding}.

\begin{figure}
    \centering
    \includegraphics[trim=0cm 14cm 0cm 0cm, clip, width=0.48\textwidth]{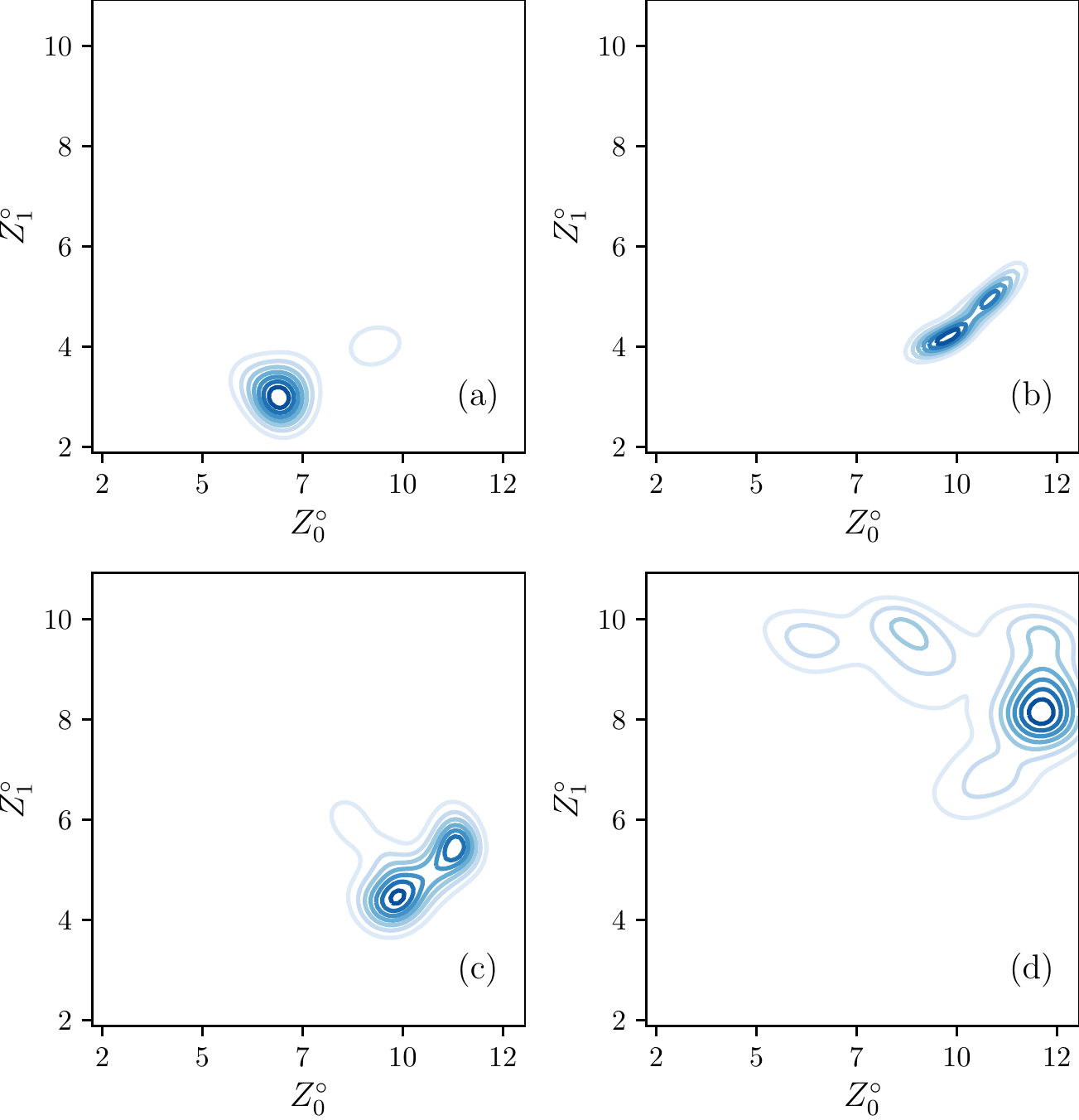}
    \includegraphics[trim=0cm 0cm 0cm 14cm, clip, width=0.48\textwidth]{figures/key-structures-histograms.pdf}
    \caption{Sample global features (i.e., $\mathbf{H}$) corresponding to the structures in Fig.~2 in the main text.
    }
    \label{fig:key-histograms}
\end{figure}

Once the feature vectors for each particle have been \textit{pooled} and \textit{embedded}, the collection of $Z^\circ$ vectors are further \textit{pooled} into a single histogram representing the entire snapshot.
This is indicated in Fig.~1 as $\textbf{H}$.
Examples from the structures in Fig.~2 of the main text are shown in Fig.~\ref{fig:key-histograms} -- note these are the actual histograms for the entire snapshot, whereas the features shown in Fig.~\ref{fig:key-features} are representative of only a single particle in those snapshots.
To complete the procedure, this $\textbf{H}$ is finally \textit{embedded} into a global structural space to yield a 2D vector which describes the global morphology.

\pagebreak
\section{Expert-knowledge-derived order parameters}

We show many order parameters for global and single chain measures, motivated by Ref.~\citenum{Statt2020} and a variety of sources from literature.\cite{rawat2011shape, dima2004asymmetry, vymetal2011gyration}

\begin{figure}[h]
    \centering
    \includegraphics[width=\textwidth]{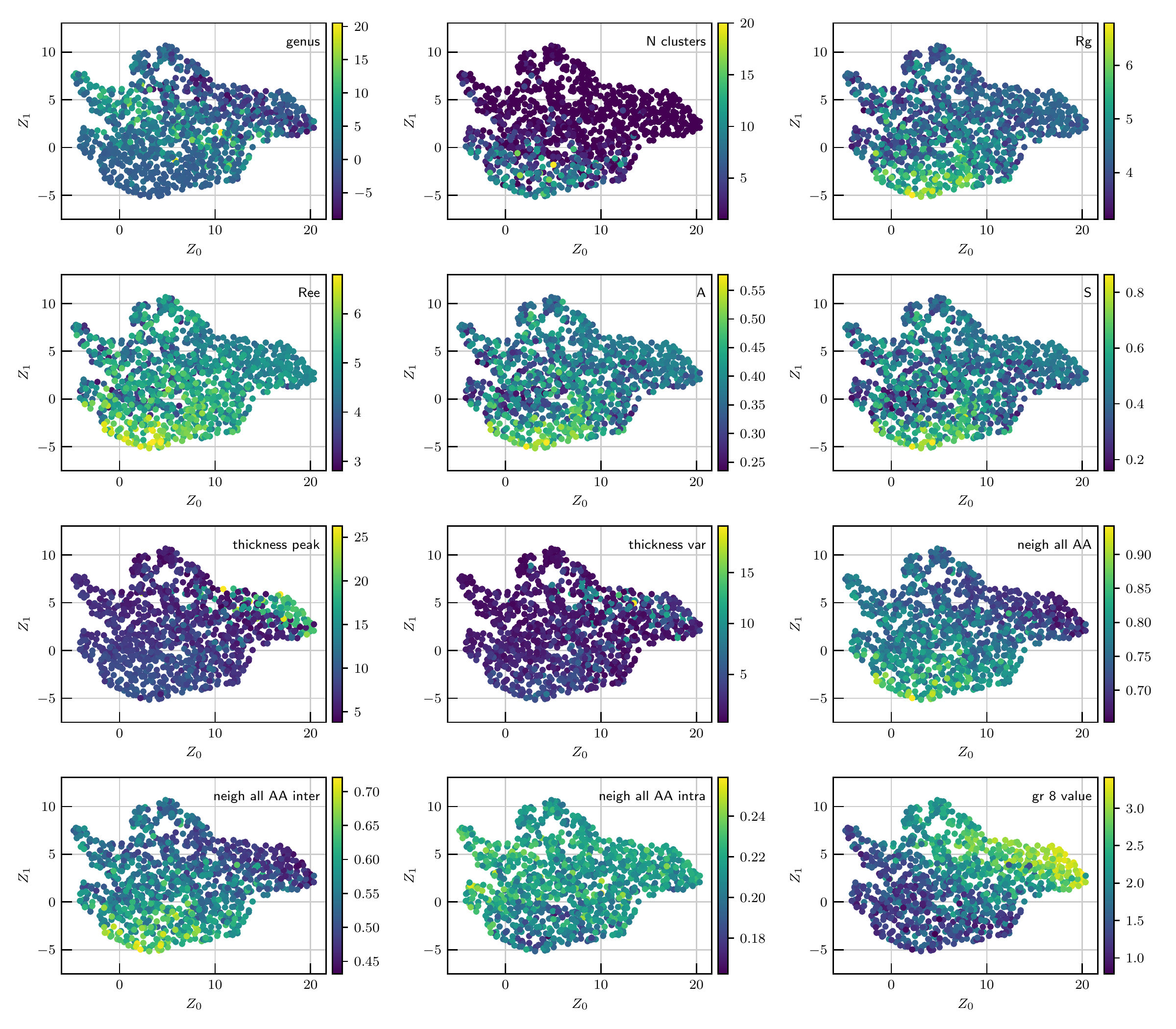}
    \caption{Data for the k-means $N=500$ manifold colored according to a variety of order parameters derived from conventional thermodynamic values.
    }
    \label{fig:order-parameters}
\end{figure}

\begin{itemize}
\item \textbf{genus}: genus of the surface (determined by Ovito, respects periodic boundary conditions, not normalized) 
\item \textbf{area}: area of the surface (determined by Ovito, respects periodic boundary conditions, not normalized) 
\item \textbf{volume}: volume of the surface (determined by Ovito, respects periodic boundary conditions, not normalized) 
\item \textbf{N clusters}: number of clusters in the system, cutoff = $1.3 \, \sigma$  for clustering 
\item \textbf{Rg}: average radius of gyration of each single chain
\item \textbf{Ree}: average end-to-end distance of each single chain
\item \textbf{A}: anisotropy value of each chain 
\item \textbf{S}: sphericity value of each chain 
\item \textbf{thickness peak}: peak location of the histogram of all thicknesses measured (obtained by fitting a Gaussian)
\item \textbf{thickness var}: peak variance of the histogram of all thicknesses measured (obtained by fitting a Gaussian)
\item \textbf{neigh all AA}: normalized number of A-neighbors in a $2 \, \sigma$ radius around each A particle
\item \textbf{neigh all AA inter}: normalized number of inter-chain A-neighbors in a $2 \, \sigma$ radius around each A particle
\item \textbf{neigh all AA intra}: normalized number of intra-chain A-neighbors in a $2 \, \sigma$ radius around each A particle
\item \textbf{gr 8 value}: value of the pair correlation function at $r = 8 \, \sigma$
\end{itemize}

For \textbf{A} and \textbf{S}:
$A$ and $S$ obey the inequalities $0 < A \le 1$ and $-\frac{1}{4} < S \le 2$, where
\begin{itemize}
\item $S < 0$: oblate
\item $S > 0$: prolate
\item $S=2, A=1$: rod shape
\item $S=0, A=0$: perfect sphere
\item $S=-\frac{1}{4}, A=1$: pancake shape
\end{itemize}

For neighbor calculation:
\begin{itemize}
    \item normalized by total number of neighbors 
    \item directly bonded neighbors do not count (since they don't interact with LJ) 
    \item should correlate well with LJ energy
\end{itemize}

For Ovito surface determination:
\begin{itemize}
    \item surface, volume, genus, and area are determined for each cluster in the system, quantities are averaged over clusters 
    \item for surface algorithm: smoothing level = 10, probe sphere radius is $2 \, \sigma$ (using \texttt{ConstructSurfaceModifier.Method.AlphaShape})
    \item for clustering: cutoff is $1.3 \, \sigma$ (using \texttt{ClusterAnalysisModifier})
    \item has unclear input parameters (probe sphere radius and smoothing level) cutoff for cluster is estimated from first non-bonded pair-correlation minimum 
\end{itemize}

For thickness calculations:
\begin{itemize}
    \item determine normal of all triangles from Ovito surface, determine distance to intersection with surface from normals
    \item fit Gaussian to most prominent peak in histogram of all intersection distances
\end{itemize}

%


\begin{thebibliography}{54}%
\makeatletter
\providecommand \@ifxundefined [1]{%
 \@ifx{#1\undefined}
}%
\providecommand \@ifnum [1]{%
 \ifnum #1\expandafter \@firstoftwo
 \else \expandafter \@secondoftwo
 \fi
}%
\providecommand \@ifx [1]{%
 \ifx #1\expandafter \@firstoftwo
 \else \expandafter \@secondoftwo
 \fi
}%
\providecommand \natexlab [1]{#1}%
\providecommand \enquote  [1]{``#1''}%
\providecommand \bibnamefont  [1]{#1}%
\providecommand \bibfnamefont [1]{#1}%
\providecommand \citenamefont [1]{#1}%
\providecommand \href@noop [0]{\@secondoftwo}%
\providecommand \href [0]{\begingroup \@sanitize@url \@href}%
\providecommand \@href[1]{\@@startlink{#1}\@@href}%
\providecommand \@@href[1]{\endgroup#1\@@endlink}%
\providecommand \@sanitize@url [0]{\catcode `\\12\catcode `\$12\catcode
  `\&12\catcode `\#12\catcode `\^12\catcode `\_12\catcode `\%12\relax}%
\providecommand \@@startlink[1]{}%
\providecommand \@@endlink[0]{}%
\providecommand \url  [0]{\begingroup\@sanitize@url \@url }%
\providecommand \@url [1]{\endgroup\@href {#1}{\urlprefix }}%
\providecommand \urlprefix  [0]{URL }%
\providecommand \Eprint [0]{\href }%
\providecommand \doibase [0]{https://doi.org/}%
\providecommand \selectlanguage [0]{\@gobble}%
\providecommand \bibinfo  [0]{\@secondoftwo}%
\providecommand \bibfield  [0]{\@secondoftwo}%
\providecommand \translation [1]{[#1]}%
\providecommand \BibitemOpen [0]{}%
\providecommand \bibitemStop [0]{}%
\providecommand \bibitemNoStop [0]{.\EOS\space}%
\providecommand \EOS [0]{\spacefactor3000\relax}%
\providecommand \BibitemShut  [1]{\csname bibitem#1\endcsname}%
\let\auto@bib@innerbib\@empty
\bibitem [{\citenamefont {Reinhart}(2021)}]{Reinhart2021}%
  \BibitemOpen
  \bibfield  {author} {\bibinfo {author} {\bibfnamefont {W.~F.}\ \bibnamefont
  {Reinhart}},\ }\bibfield  {title} {\enquote {\bibinfo {title} {Unsupervised
  learning of atomic environments from simple features},}\ }\href@noop {}
  {\bibfield  {journal} {\bibinfo  {journal} {Computational Materials Science}\
  }\textbf {\bibinfo {volume} {196}},\ \bibinfo {pages} {110511} (\bibinfo
  {year} {2021})}\BibitemShut {NoStop}%
\bibitem [{\citenamefont {Statt}\ \emph {et~al.}(2020)\citenamefont {Statt},
  \citenamefont {Casademunt}, \citenamefont {Brangwynne},\ and\ \citenamefont
  {Panagiotopoulos}}]{Statt2020}%
  \BibitemOpen
  \bibfield  {author} {\bibinfo {author} {\bibfnamefont {A.}~\bibnamefont
  {Statt}}, \bibinfo {author} {\bibfnamefont {H.}~\bibnamefont {Casademunt}},
  \bibinfo {author} {\bibfnamefont {C.~P.}\ \bibnamefont {Brangwynne}},\ and\
  \bibinfo {author} {\bibfnamefont {A.~Z.}\ \bibnamefont {Panagiotopoulos}},\
  }\bibfield  {title} {\enquote {\bibinfo {title} {Model for disordered
  proteins with strongly sequence-dependent liquid phase behavior},}\ }\href
  {https://doi.org/10.1063/1.5141095} {\bibfield  {journal} {\bibinfo
  {journal} {The Journal of Chemical Physics}\ }\textbf {\bibinfo {volume}
  {152}},\ \bibinfo {pages} {075101} (\bibinfo {year} {2020})},\ \Eprint
  {https://arxiv.org/abs/https://doi.org/10.1063/1.5141095}
  {https://doi.org/10.1063/1.5141095} \BibitemShut {NoStop}%
\bibitem [{\citenamefont {Webb}\ \emph {et~al.}(2020)\citenamefont {Webb},
  \citenamefont {Jackson}, \citenamefont {Gil},\ and\ \citenamefont
  {de~Pablo}}]{Webb2020}%
  \BibitemOpen
  \bibfield  {author} {\bibinfo {author} {\bibfnamefont {M.~A.}\ \bibnamefont
  {Webb}}, \bibinfo {author} {\bibfnamefont {N.~E.}\ \bibnamefont {Jackson}},
  \bibinfo {author} {\bibfnamefont {P.~S.}\ \bibnamefont {Gil}},\ and\ \bibinfo
  {author} {\bibfnamefont {J.~J.}\ \bibnamefont {de~Pablo}},\ }\bibfield
  {title} {\enquote {\bibinfo {title} {Targeted sequence design within the
  coarse-grained polymer genome},}\ }\href
  {https://doi.org/10.1126/sciadv.abc6216} {\bibfield  {journal} {\bibinfo
  {journal} {Science Advances}\ }\textbf {\bibinfo {volume} {6}},\ \bibinfo
  {pages} {eabc6216} (\bibinfo {year} {2020})},\ \Eprint
  {https://arxiv.org/abs/https://advances.sciencemag.org/content/6/43/eabc6216.full.pdf}
  {https://advances.sciencemag.org/content/6/43/eabc6216.full.pdf} \BibitemShut
  {NoStop}%
\bibitem [{\citenamefont {Jablonka}\ \emph {et~al.}(2021)\citenamefont
  {Jablonka}, \citenamefont {Jothiappan}, \citenamefont {Wang}, \citenamefont
  {Smit},\ and\ \citenamefont {Yoo}}]{Jablonka2021}%
  \BibitemOpen
  \bibfield  {author} {\bibinfo {author} {\bibfnamefont {K.~M.}\ \bibnamefont
  {Jablonka}}, \bibinfo {author} {\bibfnamefont {G.~M.}\ \bibnamefont
  {Jothiappan}}, \bibinfo {author} {\bibfnamefont {S.}~\bibnamefont {Wang}},
  \bibinfo {author} {\bibfnamefont {B.}~\bibnamefont {Smit}},\ and\ \bibinfo
  {author} {\bibfnamefont {B.}~\bibnamefont {Yoo}},\ }\bibfield  {title}
  {\enquote {\bibinfo {title} {Bias free multiobjective active learning for
  materials design and discovery},}\ }\href@noop {} {\bibfield  {journal}
  {\bibinfo  {journal} {Nature Communications}\ }\textbf {\bibinfo {volume}
  {12}},\ \bibinfo {pages} {1--10} (\bibinfo {year} {2021})}\BibitemShut
  {NoStop}%
\bibitem [{\citenamefont {Afzal}\ \emph {et~al.}(2019)\citenamefont {Afzal},
  \citenamefont {Haghighatlari}, \citenamefont {Ganesh}, \citenamefont
  {Cheng},\ and\ \citenamefont {Hachmann}}]{Afzal2019}%
  \BibitemOpen
  \bibfield  {author} {\bibinfo {author} {\bibfnamefont {M.~A.~F.}\
  \bibnamefont {Afzal}}, \bibinfo {author} {\bibfnamefont {M.}~\bibnamefont
  {Haghighatlari}}, \bibinfo {author} {\bibfnamefont {S.~P.}\ \bibnamefont
  {Ganesh}}, \bibinfo {author} {\bibfnamefont {C.}~\bibnamefont {Cheng}},\ and\
  \bibinfo {author} {\bibfnamefont {J.}~\bibnamefont {Hachmann}},\ }\bibfield
  {title} {\enquote {\bibinfo {title} {Accelerated discovery of
  high-refractive-index polyimides via first-principles molecular modeling,
  virtual high-throughput screening, and data mining},}\ }\href
  {https://doi.org/10.1021/acs.jpcc.9b01147} {\bibfield  {journal} {\bibinfo
  {journal} {The Journal of Physical Chemistry C}\ }\textbf {\bibinfo {volume}
  {123}},\ \bibinfo {pages} {14610--14618} (\bibinfo {year} {2019})},\ \Eprint
  {https://arxiv.org/abs/https://doi.org/10.1021/acs.jpcc.9b01147}
  {https://doi.org/10.1021/acs.jpcc.9b01147} \BibitemShut {NoStop}%
\bibitem [{\citenamefont {Shmilovich}\ \emph {et~al.}(2020)\citenamefont
  {Shmilovich}, \citenamefont {Mansbach}, \citenamefont {Sidky}, \citenamefont
  {Dunne}, \citenamefont {Panda}, \citenamefont {Tovar},\ and\ \citenamefont
  {Ferguson}}]{Shmilovich2020}%
  \BibitemOpen
  \bibfield  {author} {\bibinfo {author} {\bibfnamefont {K.}~\bibnamefont
  {Shmilovich}}, \bibinfo {author} {\bibfnamefont {R.~A.}\ \bibnamefont
  {Mansbach}}, \bibinfo {author} {\bibfnamefont {H.}~\bibnamefont {Sidky}},
  \bibinfo {author} {\bibfnamefont {O.~E.}\ \bibnamefont {Dunne}}, \bibinfo
  {author} {\bibfnamefont {S.~S.}\ \bibnamefont {Panda}}, \bibinfo {author}
  {\bibfnamefont {J.~D.}\ \bibnamefont {Tovar}},\ and\ \bibinfo {author}
  {\bibfnamefont {A.~L.}\ \bibnamefont {Ferguson}},\ }\bibfield  {title}
  {\enquote {\bibinfo {title} {Discovery of self-assembling $\pi$-conjugated
  peptides by active learning-directed coarse-grained molecular simulation},}\
  }\href@noop {} {\bibfield  {journal} {\bibinfo  {journal} {The Journal of
  Physical Chemistry B}\ }\textbf {\bibinfo {volume} {124}},\ \bibinfo {pages}
  {3873--3891} (\bibinfo {year} {2020})}\BibitemShut {NoStop}%
\bibitem [{\citenamefont {Wang}\ \emph {et~al.}(2020)\citenamefont {Wang},
  \citenamefont {Xie}, \citenamefont {France-Lanord}, \citenamefont {Berkley},
  \citenamefont {Johnson}, \citenamefont {Shao-Horn},\ and\ \citenamefont
  {Grossman}}]{Wang2020}%
  \BibitemOpen
  \bibfield  {author} {\bibinfo {author} {\bibfnamefont {Y.}~\bibnamefont
  {Wang}}, \bibinfo {author} {\bibfnamefont {T.}~\bibnamefont {Xie}}, \bibinfo
  {author} {\bibfnamefont {A.}~\bibnamefont {France-Lanord}}, \bibinfo {author}
  {\bibfnamefont {A.}~\bibnamefont {Berkley}}, \bibinfo {author} {\bibfnamefont
  {J.~A.}\ \bibnamefont {Johnson}}, \bibinfo {author} {\bibfnamefont
  {Y.}~\bibnamefont {Shao-Horn}},\ and\ \bibinfo {author} {\bibfnamefont
  {J.~C.}\ \bibnamefont {Grossman}},\ }\bibfield  {title} {\enquote {\bibinfo
  {title} {Toward designing highly conductive polymer electrolytes by machine
  learning assisted coarse-grained molecular dynamics},}\ }\href@noop {}
  {\bibfield  {journal} {\bibinfo  {journal} {Chemistry of Materials}\ }\textbf
  {\bibinfo {volume} {32}},\ \bibinfo {pages} {4144--4151} (\bibinfo {year}
  {2020})}\BibitemShut {NoStop}%
\bibitem [{\citenamefont {Matsen}(2012)}]{Matsen2012}%
  \BibitemOpen
  \bibfield  {author} {\bibinfo {author} {\bibfnamefont {M.~W.}\ \bibnamefont
  {Matsen}},\ }\bibfield  {title} {\enquote {\bibinfo {title} {Effect of
  architecture on the phase behavior of ab-type block copolymer melts},}\
  }\href {https://doi.org/10.1021/ma202782s} {\bibfield  {journal} {\bibinfo
  {journal} {Macromolecules}\ }\textbf {\bibinfo {volume} {45}},\ \bibinfo
  {pages} {2161--2165} (\bibinfo {year} {2012})}\BibitemShut {NoStop}%
\bibitem [{\citenamefont {Zhang}\ \emph {et~al.}(2017)\citenamefont {Zhang},
  \citenamefont {Deubler}, \citenamefont {Hartlieb}, \citenamefont {Martin},
  \citenamefont {Tanaka}, \citenamefont {Patyukova}, \citenamefont {Topham},
  \citenamefont {Schacher},\ and\ \citenamefont {Perrier}}]{Zhang2017}%
  \BibitemOpen
  \bibfield  {author} {\bibinfo {author} {\bibfnamefont {J.}~\bibnamefont
  {Zhang}}, \bibinfo {author} {\bibfnamefont {R.}~\bibnamefont {Deubler}},
  \bibinfo {author} {\bibfnamefont {M.}~\bibnamefont {Hartlieb}}, \bibinfo
  {author} {\bibfnamefont {L.}~\bibnamefont {Martin}}, \bibinfo {author}
  {\bibfnamefont {J.}~\bibnamefont {Tanaka}}, \bibinfo {author} {\bibfnamefont
  {E.}~\bibnamefont {Patyukova}}, \bibinfo {author} {\bibfnamefont {P.~D.}\
  \bibnamefont {Topham}}, \bibinfo {author} {\bibfnamefont {F.~H.}\
  \bibnamefont {Schacher}},\ and\ \bibinfo {author} {\bibfnamefont
  {S.}~\bibnamefont {Perrier}},\ }\bibfield  {title} {\enquote {\bibinfo
  {title} {Evolution of microphase separation with variations of segments of
  sequence-controlled multiblock copolymers},}\ }\href
  {https://doi.org/10.1021/acs.macromol.7b01831} {\bibfield  {journal}
  {\bibinfo  {journal} {Macromolecules}\ }\textbf {\bibinfo {volume} {50}},\
  \bibinfo {pages} {7380--7387} (\bibinfo {year} {2017})}\BibitemShut {NoStop}%
\bibitem [{\citenamefont {Bates}\ and\ \citenamefont
  {Bates}(2017)}]{Bates2017}%
  \BibitemOpen
  \bibfield  {author} {\bibinfo {author} {\bibfnamefont {C.~M.}\ \bibnamefont
  {Bates}}\ and\ \bibinfo {author} {\bibfnamefont {F.~S.}\ \bibnamefont
  {Bates}},\ }\bibfield  {title} {\enquote {\bibinfo {title} {50th anniversary
  perspective: Block polymers—pure potential},}\ }\href
  {https://doi.org/10.1021/acs.macromol.6b02355} {\bibfield  {journal}
  {\bibinfo  {journal} {Macromolecules}\ }\textbf {\bibinfo {volume} {50}},\
  \bibinfo {pages} {3--22} (\bibinfo {year} {2017})}\BibitemShut {NoStop}%
\bibitem [{\citenamefont {Levine}\ \emph {et~al.}(2016)\citenamefont {Levine},
  \citenamefont {Seo}, \citenamefont {Brown},\ and\ \citenamefont
  {Hall}}]{Levine2016}%
  \BibitemOpen
  \bibfield  {author} {\bibinfo {author} {\bibfnamefont {W.~G.}\ \bibnamefont
  {Levine}}, \bibinfo {author} {\bibfnamefont {Y.}~\bibnamefont {Seo}},
  \bibinfo {author} {\bibfnamefont {J.~R.}\ \bibnamefont {Brown}},\ and\
  \bibinfo {author} {\bibfnamefont {L.~M.}\ \bibnamefont {Hall}},\ }\bibfield
  {title} {\enquote {\bibinfo {title} {Effect of sequence dispersity on
  morphology of tapered diblock copolymers from molecular dynamics
  simulations},}\ }\href {https://doi.org/10.1063/1.4972141} {\bibfield
  {journal} {\bibinfo  {journal} {The Journal of Chemical Physics}\ }\textbf
  {\bibinfo {volume} {145}},\ \bibinfo {pages} {234907} (\bibinfo {year}
  {2016})}\BibitemShut {NoStop}%
\bibitem [{\citenamefont {Mai}\ and\ \citenamefont
  {Eisenberg}(2012)}]{Mai2012}%
  \BibitemOpen
  \bibfield  {author} {\bibinfo {author} {\bibfnamefont {Y.}~\bibnamefont
  {Mai}}\ and\ \bibinfo {author} {\bibfnamefont {A.}~\bibnamefont
  {Eisenberg}},\ }\bibfield  {title} {\enquote {\bibinfo {title} {Self-assembly
  of block copolymers},}\ }\href {https://doi.org/10.1039/C2CS35115C}
  {\bibfield  {journal} {\bibinfo  {journal} {Chem. Soc. Rev.}\ }\textbf
  {\bibinfo {volume} {41}},\ \bibinfo {pages} {5969--5985} (\bibinfo {year}
  {2012})}\BibitemShut {NoStop}%
\bibitem [{\citenamefont {Bates}\ \emph {et~al.}(2012)\citenamefont {Bates},
  \citenamefont {Hillmyer}, \citenamefont {Lodge}, \citenamefont {Bates},
  \citenamefont {Delaney},\ and\ \citenamefont {Fredrickson}}]{Bates2012}%
  \BibitemOpen
  \bibfield  {author} {\bibinfo {author} {\bibfnamefont {F.~S.}\ \bibnamefont
  {Bates}}, \bibinfo {author} {\bibfnamefont {M.~A.}\ \bibnamefont {Hillmyer}},
  \bibinfo {author} {\bibfnamefont {T.~P.}\ \bibnamefont {Lodge}}, \bibinfo
  {author} {\bibfnamefont {C.~M.}\ \bibnamefont {Bates}}, \bibinfo {author}
  {\bibfnamefont {K.~T.}\ \bibnamefont {Delaney}},\ and\ \bibinfo {author}
  {\bibfnamefont {G.~H.}\ \bibnamefont {Fredrickson}},\ }\bibfield  {title}
  {\enquote {\bibinfo {title} {Multiblock polymers: panacea or pandora’s
  box?}}\ }\href@noop {} {\bibfield  {journal} {\bibinfo  {journal} {Science}\
  }\textbf {\bibinfo {volume} {336}},\ \bibinfo {pages} {434--440} (\bibinfo
  {year} {2012})}\BibitemShut {NoStop}%
\bibitem [{\citenamefont {Wu}\ \emph {et~al.}(2004)\citenamefont {Wu},
  \citenamefont {Cochran}, \citenamefont {Lodge},\ and\ \citenamefont
  {Bates}}]{Wu2004}%
  \BibitemOpen
  \bibfield  {author} {\bibinfo {author} {\bibfnamefont {L.}~\bibnamefont
  {Wu}}, \bibinfo {author} {\bibfnamefont {E.~W.}\ \bibnamefont {Cochran}},
  \bibinfo {author} {\bibfnamefont {T.~P.}\ \bibnamefont {Lodge}},\ and\
  \bibinfo {author} {\bibfnamefont {F.~S.}\ \bibnamefont {Bates}},\ }\bibfield
  {title} {\enquote {\bibinfo {title} {Consequences of block number on the
  order-disorder transition and viscoelastic properties of linear (ab)n
  multiblock copolymers},}\ }\href {https://doi.org/10.1021/ma035583m}
  {\bibfield  {journal} {\bibinfo  {journal} {Macromolecules}\ }\textbf
  {\bibinfo {volume} {37}},\ \bibinfo {pages} {3360--3368} (\bibinfo {year}
  {2004})}\BibitemShut {NoStop}%
\bibitem [{\citenamefont {Pakula}\ and\ \citenamefont
  {Matyjaszewski}(1996)}]{Pakula1996}%
  \BibitemOpen
  \bibfield  {author} {\bibinfo {author} {\bibfnamefont {T.}~\bibnamefont
  {Pakula}}\ and\ \bibinfo {author} {\bibfnamefont {K.}~\bibnamefont
  {Matyjaszewski}},\ }\bibfield  {title} {\enquote {\bibinfo {title}
  {Copolymers with controlled distribution of comonomers along the chain, 1.
  structure, thermodynamics and dynamic properties of gradient copolymers.
  computer simulation},}\ }\href {https://doi.org/10.1002/mats.1996.040050514}
  {\bibfield  {journal} {\bibinfo  {journal} {Macromolecular Theory and
  Simulations}\ }\textbf {\bibinfo {volume} {5}},\ \bibinfo {pages} {987--1006}
  (\bibinfo {year} {1996})}\BibitemShut {NoStop}%
\bibitem [{\citenamefont {Beránek}, \citenamefont {Posocco},\ and\
  \citenamefont {Posel}(2020)}]{Beranek2020}%
  \BibitemOpen
  \bibfield  {author} {\bibinfo {author} {\bibfnamefont {P.}~\bibnamefont
  {Beránek}}, \bibinfo {author} {\bibfnamefont {P.}~\bibnamefont {Posocco}},\
  and\ \bibinfo {author} {\bibfnamefont {Z.}~\bibnamefont {Posel}},\ }\bibfield
   {title} {\enquote {\bibinfo {title} {Phase behavior of gradient copolymer
  melts with different gradient strengths revealed by mesoscale simulations},}\
  }\href {https://doi.org/10.3390/polym12112462} {\bibfield  {journal}
  {\bibinfo  {journal} {Polymers}\ }\textbf {\bibinfo {volume} {12}},\ \bibinfo
  {pages} {2462} (\bibinfo {year} {2020})}\BibitemShut {NoStop}%
\bibitem [{\citenamefont {Koch}\ \emph {et~al.}(2015)\citenamefont {Koch},
  \citenamefont {Panagiotopoulos}, \citenamefont {Verso},\ and\ \citenamefont
  {Likos}}]{Koch2015}%
  \BibitemOpen
  \bibfield  {author} {\bibinfo {author} {\bibfnamefont {C.}~\bibnamefont
  {Koch}}, \bibinfo {author} {\bibfnamefont {A.~Z.}\ \bibnamefont
  {Panagiotopoulos}}, \bibinfo {author} {\bibfnamefont {F.~L.}\ \bibnamefont
  {Verso}},\ and\ \bibinfo {author} {\bibfnamefont {C.~N.}\ \bibnamefont
  {Likos}},\ }\bibfield  {title} {\enquote {\bibinfo {title} {Customizing
  wormlike mesoscale structures via self-assembly of amphiphilic star
  polymers},}\ }\href@noop {} {\bibfield  {journal} {\bibinfo  {journal} {Soft
  Matter}\ }\textbf {\bibinfo {volume} {11}},\ \bibinfo {pages} {3530--3535}
  (\bibinfo {year} {2015})}\BibitemShut {NoStop}%
\bibitem [{\citenamefont {Floriano}, \citenamefont {Caponetti},\ and\
  \citenamefont {Panagiotopoulos}(1999)}]{Floriano1999}%
  \BibitemOpen
  \bibfield  {author} {\bibinfo {author} {\bibfnamefont {M.~A.}\ \bibnamefont
  {Floriano}}, \bibinfo {author} {\bibfnamefont {E.}~\bibnamefont
  {Caponetti}},\ and\ \bibinfo {author} {\bibfnamefont {A.~Z.}\ \bibnamefont
  {Panagiotopoulos}},\ }\bibfield  {title} {\enquote {\bibinfo {title}
  {Micellization in model surfactant systems},}\ }\href
  {https://doi.org/10.1021/la9810206} {\bibfield  {journal} {\bibinfo
  {journal} {Langmuir}\ }\textbf {\bibinfo {volume} {15}},\ \bibinfo {pages}
  {3143--3151} (\bibinfo {year} {1999})}\BibitemShut {NoStop}%
\bibitem [{\citenamefont {Li}, \citenamefont {Yu},\ and\ \citenamefont
  {Zhou}(2019)}]{Li2019}%
  \BibitemOpen
  \bibfield  {author} {\bibinfo {author} {\bibfnamefont {S.}~\bibnamefont
  {Li}}, \bibinfo {author} {\bibfnamefont {C.}~\bibnamefont {Yu}},\ and\
  \bibinfo {author} {\bibfnamefont {Y.}~\bibnamefont {Zhou}},\ }\bibfield
  {title} {\enquote {\bibinfo {title} {Phase diagrams, mechanisms and unique
  characteristics of alternating-structured polymer self-assembly via
  simulations},}\ }\href {https://doi.org/10.1007/s11426-018-9360-3} {\bibfield
   {journal} {\bibinfo  {journal} {Science China Chemistry}\ }\textbf {\bibinfo
  {volume} {62}},\ \bibinfo {pages} {226--237} (\bibinfo {year}
  {2019})}\BibitemShut {NoStop}%
\bibitem [{\citenamefont {Posocco}, \citenamefont {Fermeglia},\ and\
  \citenamefont {Pricl}(2010)}]{Posocco2010}%
  \BibitemOpen
  \bibfield  {author} {\bibinfo {author} {\bibfnamefont {P.}~\bibnamefont
  {Posocco}}, \bibinfo {author} {\bibfnamefont {M.}~\bibnamefont {Fermeglia}},\
  and\ \bibinfo {author} {\bibfnamefont {S.}~\bibnamefont {Pricl}},\ }\bibfield
   {title} {\enquote {\bibinfo {title} {Morphology prediction of block
  copolymers for drug delivery by mesoscale simulations},}\ }\href
  {https://doi.org/10.1039/C0JM01301C} {\bibfield  {journal} {\bibinfo
  {journal} {J. Mater. Chem.}\ }\textbf {\bibinfo {volume} {20}},\ \bibinfo
  {pages} {7742--7753} (\bibinfo {year} {2010})}\BibitemShut {NoStop}%
\bibitem [{\citenamefont {Dolgov}\ \emph {et~al.}(2018)\citenamefont {Dolgov},
  \citenamefont {Grigor’ev}, \citenamefont {Kulebyakina}, \citenamefont
  {Razuvaeva}, \citenamefont {Gumerov}, \citenamefont {Chvalun},\ and\
  \citenamefont {Potemkin}}]{Dolgov2018}%
  \BibitemOpen
  \bibfield  {author} {\bibinfo {author} {\bibfnamefont {D.}~\bibnamefont
  {Dolgov}}, \bibinfo {author} {\bibfnamefont {T.}~\bibnamefont {Grigor’ev}},
  \bibinfo {author} {\bibfnamefont {A.}~\bibnamefont {Kulebyakina}}, \bibinfo
  {author} {\bibfnamefont {E.}~\bibnamefont {Razuvaeva}}, \bibinfo {author}
  {\bibfnamefont {R.}~\bibnamefont {Gumerov}}, \bibinfo {author} {\bibfnamefont
  {S.}~\bibnamefont {Chvalun}},\ and\ \bibinfo {author} {\bibfnamefont
  {I.}~\bibnamefont {Potemkin}},\ }\bibfield  {title} {\enquote {\bibinfo
  {title} {Aggregation in biocompatible linear block copolymers: Computer
  simulation study},}\ }\href@noop {} {\bibfield  {journal} {\bibinfo
  {journal} {Polymer Science, Series A}\ }\textbf {\bibinfo {volume} {60}},\
  \bibinfo {pages} {902--910} (\bibinfo {year} {2018})}\BibitemShut {NoStop}%
\bibitem [{\citenamefont {Fenyves}\ \emph {et~al.}(2014)\citenamefont
  {Fenyves}, \citenamefont {Schmutz}, \citenamefont {Horner}, \citenamefont
  {Bright},\ and\ \citenamefont {Rzayev}}]{Fenyves2014}%
  \BibitemOpen
  \bibfield  {author} {\bibinfo {author} {\bibfnamefont {R.~D.}\ \bibnamefont
  {Fenyves}}, \bibinfo {author} {\bibfnamefont {M.}~\bibnamefont {Schmutz}},
  \bibinfo {author} {\bibfnamefont {I.}~\bibnamefont {Horner}}, \bibinfo
  {author} {\bibfnamefont {F.}~\bibnamefont {Bright}},\ and\ \bibinfo {author}
  {\bibfnamefont {J.}~\bibnamefont {Rzayev}},\ }\bibfield  {title} {\enquote
  {\bibinfo {title} {Aqueous self-assembly of giant bottlebrush block copolymer
  surfactants as shape-tunable building blocks.}}\ }\href@noop {} {\bibfield
  {journal} {\bibinfo  {journal} {Journal of the American Chemical Society}\
  }\textbf {\bibinfo {volume} {136 21}},\ \bibinfo {pages} {7762--70} (\bibinfo
  {year} {2014})}\BibitemShut {NoStop}%
\bibitem [{\citenamefont {Gindy}, \citenamefont {Prud’homme},\ and\
  \citenamefont {Panagiotopoulos}(2008)}]{Gindy2008}%
  \BibitemOpen
  \bibfield  {author} {\bibinfo {author} {\bibfnamefont {M.~E.}\ \bibnamefont
  {Gindy}}, \bibinfo {author} {\bibfnamefont {R.~K.}\ \bibnamefont
  {Prud’homme}},\ and\ \bibinfo {author} {\bibfnamefont {A.~Z.}\ \bibnamefont
  {Panagiotopoulos}},\ }\bibfield  {title} {\enquote {\bibinfo {title} {Phase
  behavior and structure formation in linear multiblock copolymer solutions by
  monte carlo simulation},}\ }\href {https://doi.org/10.1063/1.2905231}
  {\bibfield  {journal} {\bibinfo  {journal} {The Journal of Chemical Physics}\
  }\textbf {\bibinfo {volume} {128}},\ \bibinfo {pages} {164906} (\bibinfo
  {year} {2008})}\BibitemShut {NoStop}%
\bibitem [{\citenamefont {Hugouvieux}, \citenamefont {Axelos},\ and\
  \citenamefont {Kolb}(2009)}]{Hugouvieux2009}%
  \BibitemOpen
  \bibfield  {author} {\bibinfo {author} {\bibfnamefont {V.}~\bibnamefont
  {Hugouvieux}}, \bibinfo {author} {\bibfnamefont {M.~A.~V.}\ \bibnamefont
  {Axelos}},\ and\ \bibinfo {author} {\bibfnamefont {M.}~\bibnamefont {Kolb}},\
  }\bibfield  {title} {\enquote {\bibinfo {title} {Amphiphilic multiblock
  copolymers: From intramolecular pearl necklace to layered structures},}\
  }\href {https://doi.org/10.1021/ma801337a} {\bibfield  {journal} {\bibinfo
  {journal} {Macromolecules}\ }\textbf {\bibinfo {volume} {42}},\ \bibinfo
  {pages} {392--400} (\bibinfo {year} {2009})}\BibitemShut {NoStop}%
\bibitem [{\citenamefont {Hugouvieux}, \citenamefont {Axelos},\ and\
  \citenamefont {Kolb}(2011)}]{Hugouvieux2011}%
  \BibitemOpen
  \bibfield  {author} {\bibinfo {author} {\bibfnamefont {V.}~\bibnamefont
  {Hugouvieux}}, \bibinfo {author} {\bibfnamefont {M.~A.}\ \bibnamefont
  {Axelos}},\ and\ \bibinfo {author} {\bibfnamefont {M.}~\bibnamefont {Kolb}},\
  }\bibfield  {title} {\enquote {\bibinfo {title} {Micelle formation, gelation
  and phase separation of amphiphilic multiblock copolymers},}\ }\href@noop {}
  {\bibfield  {journal} {\bibinfo  {journal} {Soft Matter}\ }\textbf {\bibinfo
  {volume} {7}},\ \bibinfo {pages} {2580--2591} (\bibinfo {year}
  {2011})}\BibitemShut {NoStop}%
\bibitem [{\citenamefont {Lechner}\ and\ \citenamefont
  {Dellago}(2008)}]{Lechner2008}%
  \BibitemOpen
  \bibfield  {author} {\bibinfo {author} {\bibfnamefont {W.}~\bibnamefont
  {Lechner}}\ and\ \bibinfo {author} {\bibfnamefont {C.}~\bibnamefont
  {Dellago}},\ }\bibfield  {title} {\enquote {\bibinfo {title} {Accurate
  determination of crystal structures based on averaged local bond order
  parameters},}\ }\href {https://doi.org/10.1063/1.2977970} {\bibfield
  {journal} {\bibinfo  {journal} {The Journal of Chemical Physics}\ }\textbf
  {\bibinfo {volume} {129}},\ \bibinfo {pages} {114707} (\bibinfo {year}
  {2008})},\ \Eprint {https://arxiv.org/abs/https://doi.org/10.1063/1.2977970}
  {https://doi.org/10.1063/1.2977970} \BibitemShut {NoStop}%
\bibitem [{\citenamefont {Steinhardt}, \citenamefont {Nelson},\ and\
  \citenamefont {Ronchetti}(1983)}]{Steinhardt1983}%
  \BibitemOpen
  \bibfield  {author} {\bibinfo {author} {\bibfnamefont {P.~J.}\ \bibnamefont
  {Steinhardt}}, \bibinfo {author} {\bibfnamefont {D.~R.}\ \bibnamefont
  {Nelson}},\ and\ \bibinfo {author} {\bibfnamefont {M.}~\bibnamefont
  {Ronchetti}},\ }\bibfield  {title} {\enquote {\bibinfo {title}
  {Bond-orientational order in liquids and glasses},}\ }\href
  {https://doi.org/10.1103/PhysRevB.28.784} {\bibfield  {journal} {\bibinfo
  {journal} {Phys. Rev. B}\ }\textbf {\bibinfo {volume} {28}},\ \bibinfo
  {pages} {784--805} (\bibinfo {year} {1983})}\BibitemShut {NoStop}%
\bibitem [{\citenamefont {Reinhart}\ \emph {et~al.}(2017)\citenamefont
  {Reinhart}, \citenamefont {Long}, \citenamefont {Howard}, \citenamefont
  {Ferguson},\ and\ \citenamefont {Panagiotopoulos}}]{Reinhart2017}%
  \BibitemOpen
  \bibfield  {author} {\bibinfo {author} {\bibfnamefont {W.~F.}\ \bibnamefont
  {Reinhart}}, \bibinfo {author} {\bibfnamefont {A.~W.}\ \bibnamefont {Long}},
  \bibinfo {author} {\bibfnamefont {M.~P.}\ \bibnamefont {Howard}}, \bibinfo
  {author} {\bibfnamefont {A.~L.}\ \bibnamefont {Ferguson}},\ and\ \bibinfo
  {author} {\bibfnamefont {A.~Z.}\ \bibnamefont {Panagiotopoulos}},\ }\bibfield
   {title} {\enquote {\bibinfo {title} {Machine learning for autonomous crystal
  structure identification},}\ }\href {https://doi.org/10.1039/C7SM00957G}
  {\bibfield  {journal} {\bibinfo  {journal} {Soft Matter}\ }\textbf {\bibinfo
  {volume} {13}},\ \bibinfo {pages} {4733--4745} (\bibinfo {year}
  {2017})}\BibitemShut {NoStop}%
\bibitem [{\citenamefont {Akcasu}, \citenamefont {Benmouna},\ and\
  \citenamefont {Han}(1980)}]{AKCASU1980866}%
  \BibitemOpen
  \bibfield  {author} {\bibinfo {author} {\bibfnamefont {A.}~\bibnamefont
  {Akcasu}}, \bibinfo {author} {\bibfnamefont {M.}~\bibnamefont {Benmouna}},\
  and\ \bibinfo {author} {\bibfnamefont {C.~C.}\ \bibnamefont {Han}},\
  }\bibfield  {title} {\enquote {\bibinfo {title} {Interpretation of dynamic
  scattering from polymer solutions},}\ }\href
  {https://doi.org/https://doi.org/10.1016/0032-3861(80)90242-6} {\bibfield
  {journal} {\bibinfo  {journal} {Polymer}\ }\textbf {\bibinfo {volume} {21}},\
  \bibinfo {pages} {866--890} (\bibinfo {year} {1980})}\BibitemShut {NoStop}%
\bibitem [{\citenamefont {Pedersen}(1997)}]{PEDERSEN1997171}%
  \BibitemOpen
  \bibfield  {author} {\bibinfo {author} {\bibfnamefont {J.~S.}\ \bibnamefont
  {Pedersen}},\ }\bibfield  {title} {\enquote {\bibinfo {title} {Analysis of
  small-angle scattering data from colloids and polymer solutions: modeling and
  least-squares fitting},}\ }\href
  {https://doi.org/https://doi.org/10.1016/S0001-8686(97)00312-6} {\bibfield
  {journal} {\bibinfo  {journal} {Advances in Colloid and Interface Science}\
  }\textbf {\bibinfo {volume} {70}},\ \bibinfo {pages} {171--210} (\bibinfo
  {year} {1997})}\BibitemShut {NoStop}%
\bibitem [{\citenamefont {Coifman}\ and\ \citenamefont
  {Lafon}(2006)}]{Coifman2006}%
  \BibitemOpen
  \bibfield  {author} {\bibinfo {author} {\bibfnamefont {R.~R.}\ \bibnamefont
  {Coifman}}\ and\ \bibinfo {author} {\bibfnamefont {S.}~\bibnamefont
  {Lafon}},\ }\bibfield  {title} {\enquote {\bibinfo {title} {Diffusion
  maps},}\ }\href@noop {} {\bibfield  {journal} {\bibinfo  {journal} {Applied
  and computational harmonic analysis}\ }\textbf {\bibinfo {volume} {21}},\
  \bibinfo {pages} {5--30} (\bibinfo {year} {2006})}\BibitemShut {NoStop}%
\bibitem [{\citenamefont {Ferguson}\ \emph {et~al.}(2011)\citenamefont
  {Ferguson}, \citenamefont {Panagiotopoulos}, \citenamefont {Kevrekidis},\
  and\ \citenamefont {Debenedetti}}]{Ferguson2011}%
  \BibitemOpen
  \bibfield  {author} {\bibinfo {author} {\bibfnamefont {A.~L.}\ \bibnamefont
  {Ferguson}}, \bibinfo {author} {\bibfnamefont {A.~Z.}\ \bibnamefont
  {Panagiotopoulos}}, \bibinfo {author} {\bibfnamefont {I.~G.}\ \bibnamefont
  {Kevrekidis}},\ and\ \bibinfo {author} {\bibfnamefont {P.~G.}\ \bibnamefont
  {Debenedetti}},\ }\bibfield  {title} {\enquote {\bibinfo {title} {Nonlinear
  dimensionality reduction in molecular simulation: The diffusion map
  approach},}\ }\href@noop {} {\bibfield  {journal} {\bibinfo  {journal}
  {Chemical Physics Letters}\ }\textbf {\bibinfo {volume} {509}},\ \bibinfo
  {pages} {1--11} (\bibinfo {year} {2011})}\BibitemShut {NoStop}%
\bibitem [{\citenamefont {Long}\ \emph {et~al.}(2015)\citenamefont {Long},
  \citenamefont {Zhang}, \citenamefont {Granick},\ and\ \citenamefont
  {Ferguson}}]{Long2015}%
  \BibitemOpen
  \bibfield  {author} {\bibinfo {author} {\bibfnamefont {A.~W.}\ \bibnamefont
  {Long}}, \bibinfo {author} {\bibfnamefont {J.}~\bibnamefont {Zhang}},
  \bibinfo {author} {\bibfnamefont {S.}~\bibnamefont {Granick}},\ and\ \bibinfo
  {author} {\bibfnamefont {A.~L.}\ \bibnamefont {Ferguson}},\ }\bibfield
  {title} {\enquote {\bibinfo {title} {Machine learning assembly landscapes
  from particle tracking data},}\ }\href@noop {} {\bibfield  {journal}
  {\bibinfo  {journal} {Soft Matter}\ }\textbf {\bibinfo {volume} {11}},\
  \bibinfo {pages} {8141--8153} (\bibinfo {year} {2015})}\BibitemShut {NoStop}%
\bibitem [{\citenamefont {Wang}\ and\ \citenamefont
  {Ferguson}(2018)}]{Wang2018}%
  \BibitemOpen
  \bibfield  {author} {\bibinfo {author} {\bibfnamefont {J.}~\bibnamefont
  {Wang}}\ and\ \bibinfo {author} {\bibfnamefont {A.~L.}\ \bibnamefont
  {Ferguson}},\ }\bibfield  {title} {\enquote {\bibinfo {title} {A study of the
  morphology, dynamics, and folding pathways of ring polymers with
  supramolecular topological constraints using molecular simulation and
  nonlinear manifold learning},}\ }\href@noop {} {\bibfield  {journal}
  {\bibinfo  {journal} {Macromolecules}\ }\textbf {\bibinfo {volume} {51}},\
  \bibinfo {pages} {598--616} (\bibinfo {year} {2018})}\BibitemShut {NoStop}%
\bibitem [{\citenamefont {Chiappini}, \citenamefont {Patti},\ and\
  \citenamefont {Dijkstra}(2020)}]{Chiappini2020}%
  \BibitemOpen
  \bibfield  {author} {\bibinfo {author} {\bibfnamefont {M.}~\bibnamefont
  {Chiappini}}, \bibinfo {author} {\bibfnamefont {A.}~\bibnamefont {Patti}},\
  and\ \bibinfo {author} {\bibfnamefont {M.}~\bibnamefont {Dijkstra}},\
  }\bibfield  {title} {\enquote {\bibinfo {title} {Helicoidal dynamics of
  biaxial curved rods in twist-bend nematic phases unveiled by unsupervised
  machine learning techniques},}\ }\href@noop {} {\bibfield  {journal}
  {\bibinfo  {journal} {Physical Review E}\ }\textbf {\bibinfo {volume}
  {102}},\ \bibinfo {pages} {040601} (\bibinfo {year} {2020})}\BibitemShut
  {NoStop}%
\bibitem [{\citenamefont {Zwanzig}(2001)}]{Zwanzig2001}%
  \BibitemOpen
  \bibfield  {author} {\bibinfo {author} {\bibfnamefont {R.}~\bibnamefont
  {Zwanzig}},\ }\href@noop {} {\emph {\bibinfo {title} {Nonequilibrium
  statistical mechanics}}}\ (\bibinfo  {publisher} {Oxford university press},\
  \bibinfo {year} {2001})\BibitemShut {NoStop}%
\bibitem [{\citenamefont {Xu}\ \emph {et~al.}(2019)\citenamefont {Xu},
  \citenamefont {Wei}, \citenamefont {Li}, \citenamefont {Wang}, \citenamefont
  {Chen},\ and\ \citenamefont {Jiang}}]{Xu2019}%
  \BibitemOpen
  \bibfield  {author} {\bibinfo {author} {\bibfnamefont {X.}~\bibnamefont
  {Xu}}, \bibinfo {author} {\bibfnamefont {Q.}~\bibnamefont {Wei}}, \bibinfo
  {author} {\bibfnamefont {H.}~\bibnamefont {Li}}, \bibinfo {author}
  {\bibfnamefont {Y.}~\bibnamefont {Wang}}, \bibinfo {author} {\bibfnamefont
  {Y.}~\bibnamefont {Chen}},\ and\ \bibinfo {author} {\bibfnamefont
  {Y.}~\bibnamefont {Jiang}},\ }\bibfield  {title} {\enquote {\bibinfo {title}
  {Recognition of polymer configurations by unsupervised learning},}\
  }\href@noop {} {\bibfield  {journal} {\bibinfo  {journal} {Physical Review
  E}\ }\textbf {\bibinfo {volume} {99}},\ \bibinfo {pages} {043307} (\bibinfo
  {year} {2019})}\BibitemShut {NoStop}%
\bibitem [{\citenamefont {Bejagam}\ \emph {et~al.}(2018)\citenamefont
  {Bejagam}, \citenamefont {An}, \citenamefont {Singh},\ and\ \citenamefont
  {Deshmukh}}]{Bejagam2018}%
  \BibitemOpen
  \bibfield  {author} {\bibinfo {author} {\bibfnamefont {K.~K.}\ \bibnamefont
  {Bejagam}}, \bibinfo {author} {\bibfnamefont {Y.}~\bibnamefont {An}},
  \bibinfo {author} {\bibfnamefont {S.}~\bibnamefont {Singh}},\ and\ \bibinfo
  {author} {\bibfnamefont {S.~A.}\ \bibnamefont {Deshmukh}},\ }\bibfield
  {title} {\enquote {\bibinfo {title} {Machine-learning enabled new insights
  into the coil-to-globule transition of thermosensitive polymers using a
  coarse-grained model},}\ }\href@noop {} {\bibfield  {journal} {\bibinfo
  {journal} {The journal of physical chemistry letters}\ }\textbf {\bibinfo
  {volume} {9}},\ \bibinfo {pages} {6480--6488} (\bibinfo {year}
  {2018})}\BibitemShut {NoStop}%
\bibitem [{\citenamefont {Ziolek}\ \emph {et~al.}(2021)\citenamefont {Ziolek},
  \citenamefont {Smith}, \citenamefont {Pink}, \citenamefont {Dreiss},\ and\
  \citenamefont {Lorenz}}]{Ziolek2021}%
  \BibitemOpen
  \bibfield  {author} {\bibinfo {author} {\bibfnamefont {R.~M.}\ \bibnamefont
  {Ziolek}}, \bibinfo {author} {\bibfnamefont {P.}~\bibnamefont {Smith}},
  \bibinfo {author} {\bibfnamefont {D.~L.}\ \bibnamefont {Pink}}, \bibinfo
  {author} {\bibfnamefont {C.~A.}\ \bibnamefont {Dreiss}},\ and\ \bibinfo
  {author} {\bibfnamefont {C.~D.}\ \bibnamefont {Lorenz}},\ }\bibfield  {title}
  {\enquote {\bibinfo {title} {Unsupervised learning unravels the structure of
  four-arm and linear block copolymer micelles},}\ }\href@noop {} {\bibfield
  {journal} {\bibinfo  {journal} {Macromolecules}\ }\textbf {\bibinfo {volume}
  {54}},\ \bibinfo {pages} {3755--3768} (\bibinfo {year} {2021})}\BibitemShut
  {NoStop}%
\bibitem [{\citenamefont {Chen}, \citenamefont {Tan},\ and\ \citenamefont
  {Ferguson}(2018)}]{Chen2018}%
  \BibitemOpen
  \bibfield  {author} {\bibinfo {author} {\bibfnamefont {W.}~\bibnamefont
  {Chen}}, \bibinfo {author} {\bibfnamefont {A.~R.}\ \bibnamefont {Tan}},\ and\
  \bibinfo {author} {\bibfnamefont {A.~L.}\ \bibnamefont {Ferguson}},\
  }\bibfield  {title} {\enquote {\bibinfo {title} {Collective variable
  discovery and enhanced sampling using autoencoders: Innovations in network
  architecture and error function design},}\ }\href@noop {} {\bibfield
  {journal} {\bibinfo  {journal} {The Journal of chemical physics}\ }\textbf
  {\bibinfo {volume} {149}},\ \bibinfo {pages} {072312} (\bibinfo {year}
  {2018})}\BibitemShut {NoStop}%
\bibitem [{\citenamefont {Sun}\ \emph {et~al.}(2020)\citenamefont {Sun},
  \citenamefont {Li}, \citenamefont {Zhang}, \citenamefont {Gao},\ and\
  \citenamefont {Luo}}]{Sun2020}%
  \BibitemOpen
  \bibfield  {author} {\bibinfo {author} {\bibfnamefont {L.-W.}\ \bibnamefont
  {Sun}}, \bibinfo {author} {\bibfnamefont {H.}~\bibnamefont {Li}}, \bibinfo
  {author} {\bibfnamefont {X.-Q.}\ \bibnamefont {Zhang}}, \bibinfo {author}
  {\bibfnamefont {H.-B.}\ \bibnamefont {Gao}},\ and\ \bibinfo {author}
  {\bibfnamefont {M.-B.}\ \bibnamefont {Luo}},\ }\bibfield  {title} {\enquote
  {\bibinfo {title} {Identifying conformation states of polymer through
  unsupervised machine learning},}\ }\href@noop {} {\bibfield  {journal}
  {\bibinfo  {journal} {Chinese Journal of Polymer Science}\ }\textbf {\bibinfo
  {volume} {38}},\ \bibinfo {pages} {1403--1408} (\bibinfo {year}
  {2020})}\BibitemShut {NoStop}%
\bibitem [{\citenamefont {Bhattacharya}\ and\ \citenamefont
  {Patra}(2021)}]{Bhattacharya2021}%
  \BibitemOpen
  \bibfield  {author} {\bibinfo {author} {\bibfnamefont {D.}~\bibnamefont
  {Bhattacharya}}\ and\ \bibinfo {author} {\bibfnamefont {T.~K.}\ \bibnamefont
  {Patra}},\ }\bibfield  {title} {\enquote {\bibinfo {title} {dpoly: Deep
  learning of polymer phases and phase transition},}\ }\href@noop {} {\bibfield
   {journal} {\bibinfo  {journal} {Macromolecules}\ }\textbf {\bibinfo {volume}
  {54}},\ \bibinfo {pages} {3065--3074} (\bibinfo {year} {2021})}\BibitemShut
  {NoStop}%
\bibitem [{\citenamefont {Jones}\ and\ \citenamefont
  {Chapman}(1924)}]{Jones1924}%
  \BibitemOpen
  \bibfield  {author} {\bibinfo {author} {\bibfnamefont {J.~E.}\ \bibnamefont
  {Jones}}\ and\ \bibinfo {author} {\bibfnamefont {S.}~\bibnamefont
  {Chapman}},\ }\bibfield  {title} {\enquote {\bibinfo {title} {On the
  determination of molecular fields. \&\#x2014;ii. from the equation of state
  of a gas},}\ }\href {https://doi.org/10.1098/rspa.1924.0082} {\bibfield
  {journal} {\bibinfo  {journal} {Proceedings of the Royal Society of London.
  Series A, Containing Papers of a Mathematical and Physical Character}\
  }\textbf {\bibinfo {volume} {106}},\ \bibinfo {pages} {463--477} (\bibinfo
  {year} {1924})},\ \Eprint
  {https://arxiv.org/abs/https://royalsocietypublishing.org/doi/pdf/10.1098/rspa.1924.0082}
  {https://royalsocietypublishing.org/doi/pdf/10.1098/rspa.1924.0082}
  \BibitemShut {NoStop}%
\bibitem [{\citenamefont {Weeks}, \citenamefont {Chandler},\ and\ \citenamefont
  {Andersen}(1971)}]{Weeks1971}%
  \BibitemOpen
  \bibfield  {author} {\bibinfo {author} {\bibfnamefont {J.~D.}\ \bibnamefont
  {Weeks}}, \bibinfo {author} {\bibfnamefont {D.}~\bibnamefont {Chandler}},\
  and\ \bibinfo {author} {\bibfnamefont {H.~C.}\ \bibnamefont {Andersen}},\
  }\bibfield  {title} {\enquote {\bibinfo {title} {Role of repulsive forces in
  determining the equilibrium structure of simple liquids},}\ }\href
  {https://doi.org/10.1063/1.1674820} {\bibfield  {journal} {\bibinfo
  {journal} {The Journal of Chemical Physics}\ }\textbf {\bibinfo {volume}
  {54}},\ \bibinfo {pages} {5237--5247} (\bibinfo {year} {1971})}\BibitemShut
  {NoStop}%
\bibitem [{\citenamefont {Kremer}\ and\ \citenamefont
  {Grest}(1990)}]{Kremer1990}%
  \BibitemOpen
  \bibfield  {author} {\bibinfo {author} {\bibfnamefont {K.}~\bibnamefont
  {Kremer}}\ and\ \bibinfo {author} {\bibfnamefont {G.~S.}\ \bibnamefont
  {Grest}},\ }\bibfield  {title} {\enquote {\bibinfo {title} {Dynamics of
  entangled linear polymer melts: A molecular‐dynamics simulation},}\ }\href
  {https://doi.org/10.1063/1.458541} {\bibfield  {journal} {\bibinfo  {journal}
  {The Journal of Chemical Physics}\ }\textbf {\bibinfo {volume} {92}},\
  \bibinfo {pages} {5057--5086} (\bibinfo {year} {1990})},\ \Eprint
  {https://arxiv.org/abs/https://doi.org/10.1063/1.458541}
  {https://doi.org/10.1063/1.458541} \BibitemShut {NoStop}%
\bibitem [{\citenamefont {Glaser}\ \emph {et~al.}(2015)\citenamefont {Glaser},
  \citenamefont {Nguyen}, \citenamefont {Anderson}, \citenamefont {Lui},
  \citenamefont {Spiga}, \citenamefont {Millan}, \citenamefont {Morse},\ and\
  \citenamefont {Glotzer}}]{Glaser2015}%
  \BibitemOpen
  \bibfield  {author} {\bibinfo {author} {\bibfnamefont {J.}~\bibnamefont
  {Glaser}}, \bibinfo {author} {\bibfnamefont {T.~D.}\ \bibnamefont {Nguyen}},
  \bibinfo {author} {\bibfnamefont {J.~A.}\ \bibnamefont {Anderson}}, \bibinfo
  {author} {\bibfnamefont {P.}~\bibnamefont {Lui}}, \bibinfo {author}
  {\bibfnamefont {F.}~\bibnamefont {Spiga}}, \bibinfo {author} {\bibfnamefont
  {J.~A.}\ \bibnamefont {Millan}}, \bibinfo {author} {\bibfnamefont {D.~C.}\
  \bibnamefont {Morse}},\ and\ \bibinfo {author} {\bibfnamefont {S.~C.}\
  \bibnamefont {Glotzer}},\ }\bibfield  {title} {\enquote {\bibinfo {title}
  {Strong scaling of general-purpose molecular dynamics simulations on gpus},}\
  }\href {https://doi.org/https://doi.org/10.1016/j.cpc.2015.02.028} {\bibfield
   {journal} {\bibinfo  {journal} {Computer Physics Communications}\ }\textbf
  {\bibinfo {volume} {192}},\ \bibinfo {pages} {97 -- 107} (\bibinfo {year}
  {2015})}\BibitemShut {NoStop}%
\bibitem [{\citenamefont {Anderson}, \citenamefont {Lorenz},\ and\
  \citenamefont {Travesset}(2008)}]{Anderson2008}%
  \BibitemOpen
  \bibfield  {author} {\bibinfo {author} {\bibfnamefont {J.~A.}\ \bibnamefont
  {Anderson}}, \bibinfo {author} {\bibfnamefont {C.~D.}\ \bibnamefont
  {Lorenz}},\ and\ \bibinfo {author} {\bibfnamefont {A.}~\bibnamefont
  {Travesset}},\ }\bibfield  {title} {\enquote {\bibinfo {title} {General
  purpose molecular dynamics simulations fully implemented on graphics
  processing units},}\ }\href
  {https://doi.org/https://doi.org/10.1016/j.jcp.2008.01.047} {\bibfield
  {journal} {\bibinfo  {journal} {Journal of Computational Physics}\ }\textbf
  {\bibinfo {volume} {227}},\ \bibinfo {pages} {5342 -- 5359} (\bibinfo {year}
  {2008})}\BibitemShut {NoStop}%
\bibitem [{SI()}]{SI}%
  \BibitemOpen
  \href {https://SI-url} {\enquote {\bibinfo {title} {See https://si-url for
  supplementary material},}\ }\BibitemShut {NoStop}%
\bibitem [{\citenamefont {McInnes}, \citenamefont {Healy},\ and\ \citenamefont
  {Melville}(2020)}]{Mcinnes2020}%
  \BibitemOpen
  \bibfield  {author} {\bibinfo {author} {\bibfnamefont {L.}~\bibnamefont
  {McInnes}}, \bibinfo {author} {\bibfnamefont {J.}~\bibnamefont {Healy}},\
  and\ \bibinfo {author} {\bibfnamefont {J.}~\bibnamefont {Melville}},\
  }\href@noop {} {\enquote {\bibinfo {title} {Umap: Uniform manifold
  approximation and projection for dimension reduction},}\ } (\bibinfo {year}
  {2020}),\ \Eprint {https://arxiv.org/abs/1802.03426} {arXiv:1802.03426
  [stat.ML]} \BibitemShut {NoStop}%
\bibitem [{\citenamefont {Stukowski}(2014)}]{Stukowski2014}%
  \BibitemOpen
  \bibfield  {author} {\bibinfo {author} {\bibfnamefont {A.}~\bibnamefont
  {Stukowski}},\ }\bibfield  {title} {\enquote {\bibinfo {title} {Computational
  analysis methods in atomistic modeling of crystals},}\ }\href@noop {}
  {\bibfield  {journal} {\bibinfo  {journal} {Jom}\ }\textbf {\bibinfo {volume}
  {66}},\ \bibinfo {pages} {399--407} (\bibinfo {year} {2014})}\BibitemShut
  {NoStop}%
\bibitem [{\citenamefont {Dutta}\ \emph {et~al.}(2020)\citenamefont {Dutta},
  \citenamefont {Ke}, \citenamefont {Xi}, \citenamefont {Yin}, \citenamefont
  {Zhou},\ and\ \citenamefont {Ge}}]{Dutta2020}%
  \BibitemOpen
  \bibfield  {author} {\bibinfo {author} {\bibfnamefont {D.}~\bibnamefont
  {Dutta}}, \bibinfo {author} {\bibfnamefont {W.}~\bibnamefont {Ke}}, \bibinfo
  {author} {\bibfnamefont {L.}~\bibnamefont {Xi}}, \bibinfo {author}
  {\bibfnamefont {W.}~\bibnamefont {Yin}}, \bibinfo {author} {\bibfnamefont
  {M.}~\bibnamefont {Zhou}},\ and\ \bibinfo {author} {\bibfnamefont
  {Z.}~\bibnamefont {Ge}},\ }\bibfield  {title} {\enquote {\bibinfo {title}
  {{Block copolymer prodrugs: Synthesis, self-assembly, and applications for
  cancer therapy}},}\ }\href {https://doi.org/10.1002/wnan.1585} {\bibfield
  {journal} {\bibinfo  {journal} {Wiley Interdisciplinary Reviews: Nanomedicine
  and Nanobiotechnology}\ }\textbf {\bibinfo {volume} {12}},\ \bibinfo {pages}
  {1--19} (\bibinfo {year} {2020})}\BibitemShut {NoStop}%
\bibitem [{\citenamefont {Rawat}\ and\ \citenamefont
  {Biswas}(2011)}]{rawat2011shape}%
  \BibitemOpen
  \bibfield  {author} {\bibinfo {author} {\bibfnamefont {N.}~\bibnamefont
  {Rawat}}\ and\ \bibinfo {author} {\bibfnamefont {P.}~\bibnamefont {Biswas}},\
  }\bibfield  {title} {\enquote {\bibinfo {title} {Shape, flexibility and
  packing of proteins and nucleic acids in complexes},}\ }\href@noop {}
  {\bibfield  {journal} {\bibinfo  {journal} {Physical Chemistry Chemical
  Physics}\ }\textbf {\bibinfo {volume} {13}},\ \bibinfo {pages} {9632--9643}
  (\bibinfo {year} {2011})}\BibitemShut {NoStop}%
\bibitem [{\citenamefont {Dima}\ and\ \citenamefont
  {Thirumalai}(2004)}]{dima2004asymmetry}%
  \BibitemOpen
  \bibfield  {author} {\bibinfo {author} {\bibfnamefont {R.~I.}\ \bibnamefont
  {Dima}}\ and\ \bibinfo {author} {\bibfnamefont {D.}~\bibnamefont
  {Thirumalai}},\ }\bibfield  {title} {\enquote {\bibinfo {title} {Asymmetry in
  the shapes of folded and denatured states of proteins},}\ }\href@noop {}
  {\bibfield  {journal} {\bibinfo  {journal} {The Journal of Physical Chemistry
  B}\ }\textbf {\bibinfo {volume} {108}},\ \bibinfo {pages} {6564--6570}
  (\bibinfo {year} {2004})}\BibitemShut {NoStop}%
\bibitem [{\citenamefont {Vymetal}\ and\ \citenamefont
  {Vondrasek}(2011)}]{vymetal2011gyration}%
  \BibitemOpen
  \bibfield  {author} {\bibinfo {author} {\bibfnamefont {J.}~\bibnamefont
  {Vymetal}}\ and\ \bibinfo {author} {\bibfnamefont {J.}~\bibnamefont
  {Vondrasek}},\ }\bibfield  {title} {\enquote {\bibinfo {title} {Gyration-and
  inertia-tensor-based collective coordinates for metadynamics. application on
  the conformational behavior of polyalanine peptides and trp-cage folding},}\
  }\href@noop {} {\bibfield  {journal} {\bibinfo  {journal} {The Journal of
  Physical Chemistry A}\ }\textbf {\bibinfo {volume} {115}},\ \bibinfo {pages}
  {11455--11465} (\bibinfo {year} {2011})}\BibitemShut {NoStop}%
\end{thebibliography}
\end{document}